\RequirePackage{lineno}
\documentclass[aps,iopart,twocolumn,superscriptaddress,amsmath,amssymb]{revtex4}
\usepackage{graphicx} 
\usepackage{epstopdf} 
\usepackage{dcolumn} 
\usepackage{xcolor} 
\usepackage{amsmath,amssymb} 
\usepackage[colorlinks,
            linkcolor = blue,
            anchorcolor = blue,
            urlcolor = blue,
            citecolor=blue
]{hyperref} 
\usepackage[utf8]{inputenc} 
\usepackage[english]{babel} 
\usepackage{ifthen} 
\usepackage{orcidlink}
\usepackage{gensymb}
\usepackage{multirow}

\graphicspath{{figures/}} 

\input{belle2-symbols}

\newcolumntype{L}[1]{>{\raggedright\let\newline\\\arraybackslash\hspace{0pt}}m{#1}}
\newcolumntype{C}[1]{>{\centering\let\newline\\\arraybackslash\hspace{0pt}}m{#1}}
\newcolumntype{R}[1]{>{\raggedleft\let\newline\\\arraybackslash\hspace{0pt}}m{#1}}

\let\oldequation\equation
\let\oldendequation\endequation

\renewenvironment{equation}{\linenomathNonumbers\oldequation}{\oldendequation\endlinenomath}
\begin{document}


\title{Measurement of the integrated luminosity of data samples collected during 2019-2022 by the Belle~II experiment}

  \author{I.~Adachi\,\orcidlink{0000-0003-2287-0173}} 
  \author{L.~Aggarwal\,\orcidlink{0000-0002-0909-7537}} 
  \author{H.~Ahmed\,\orcidlink{0000-0003-3976-7498}} 
  \author{J.~K.~Ahn\,\orcidlink{0000-0002-5795-2243}} 
  \author{H.~Aihara\,\orcidlink{0000-0002-1907-5964}} 
  \author{N.~Akopov\,\orcidlink{0000-0002-4425-2096}} 
  \author{A.~Aloisio\,\orcidlink{0000-0002-3883-6693}} 
  \author{N.~Althubiti\,\orcidlink{0000-0003-1513-0409}} 
  \author{N.~Anh~Ky\,\orcidlink{0000-0003-0471-197X}} 
  \author{D.~M.~Asner\,\orcidlink{0000-0002-1586-5790}} 
% ...
\collaboration{The Belle II Collaboration}

\begin{abstract}
A series of data samples was collected with the Belle~II detector at the SuperKEKB collider from March 2019 to June 2022. We determine the integrated luminosities of these data samples using three distinct methodologies involving Bhabha ($e^+e^- \to e^+e^-(n\gamma)$), digamma ($e^+e^- \to \gamma\gamma(n\gamma)$), and dimuon ($e^+e^- \to \mu^+ \mu^- (n\gamma)$) events. The total integrated luminosity obtained with Bhabha, digamma, and dimuon events is ({426.88} $\pm$ 0.03 $\pm$ {2.61})~fb$^{-1}$, ({429.28} $\pm$ 0.03 $\pm$ {2.62})~fb$^{-1}$, and ({423.99} $\pm$ 0.04 $\pm$ {3.83})~fb$^{-1}$, where the first uncertainties are statistical and the second are systematic. The resulting total integrated luminosity obtained from the combination of the three methods is ({427.87 $\pm$ 2.01})~fb$^{-1}$.

\end{abstract}

\maketitle

\section{Introduction}

Integrated luminosity, denoted $\cal L$, connects the number of produced events ($N$) with the cross section ($\sigma$) of a specific physical process, as expressed by the formula
  \begin{equation}\label{eq:1}
    \emph{N} = {\cal L} \cdot \sigma.
  \end{equation} 
Accurate measurements of integrated luminosity are essential to minimize uncertainties in cross-section measurements of interesting physics processes.

In high-energy physics experiments, integrated luminosity is determined using Eq.~(\ref{eq:1}), employing well-known quantum electrodynamics processes. In this paper, we present the results of an integrated luminosity measurement conducted by the Belle~II experiment~\cite{Abe:2010gxa}, which operates at the SuperKEKB $e^{+}e^{-}$ collider~\cite{Akai:2018mbz} located at the High Energy Accelerator Research Organization (KEK) in Japan. SuperKEKB is an asymmetric energy collider in which electrons with an energy of 7~GeV and positrons with an energy of 4~GeV circulate in the high energy ring (HER) and low energy ring (LER), respectively. The collider mainly operates at the center-of-mass (c.m.)\ energy of 10.58~GeV, which is at the peak of the $\Upsilon(4S)$ resonance. Belle~II is designed to measure the parameters of the Standard Model precisely and search for new physics beyond the Standard Model with a planned integrated luminosity of 50~ab$^{-1}$. The instantaneous luminosity target is $6 \times 10^{35}$~${\rm cm^{-2}s^{-1}}$. In June 2022, SuperKEKB achieved an instantaneous luminosity world record of $4.7 \times 10^{34}$~${\rm cm^{-2}s^{-1}}$, with a statistical uncertainty of 2.7\% and a systematic uncertainty of 1.7\%~\cite{Kovalenko:2020ftl}. Determination of the integrated luminosity directly via integration of the instantaneous luminosity therefore carries a substantial systematic uncertainty.

The Belle~II/SuperKEKB project has had three major commissioning phases. Phase~1 was carried out in Spring 2016 without beam collisions and was prior to the installation of the Belle~II detector. Phase~2, which began in March 2018 and ended in July 2018, marked the beginning of $e^+e^-$ collisions with the Belle~II detector installed, albeit without the vertex detector (VXD). The integrated luminosity of Phase~2 was measured to be (496.3 $\pm$ 0.3 $\pm$ 3.0)~pb$^{-1}$~\cite{Phase2_lum}. Phase~3 started in March 2019 with a partial VXD and resulted in data for physics analyses. The data sample collected from March 2019 to June 2022 is defined as the Run~1 data. 

We measure the integrated luminosity of the Run~1 data using the following three precisely calculable processes: Bhabha scattering ($e^+e^- \to e^+e^-(n\gamma)$), digamma production ($e^+e^- \to \gamma\gamma(n\gamma)$), and dimuon production ($e^+e^- \to \mu^+\mu^-(n\gamma)$). These processes benefit from large and well-known production cross sections and clean experimental signatures.

Several improvements and a new cross-check have been implemented with respect to the method developed for the Phase~2 dataset. In the Run~1 data, along with the installation of the VXD detector system in the Belle~II detector, improvements were implemented in the reconstruction procedure. These improvements enable the use of tracks to select Bhabha events and implementation of a new cross-check with dimuon events, reducing the dependence on the electromagnetic calorimeter~(ECL) cluster information. The Run~1 data sample has almost 1,000 times the integrated luminosity and 10 times the instantaneous luminosity of that in Phase~2. However, higher instantaneous luminosities result in increased beam-induced background~\cite{beamBKG}. A new analysis method has been developed to assess the impact of higher background levels. To address the larger background levels, a comprehensive trigger system is implemented consisting of both a hardware-based level~1 trigger (L1 trigger)~\cite{L1Trigger,L1Trigger2} and a software-based high-level trigger (HLT)~\cite{hlt} (Section~\ref{sec:corrections}).

\section{The Belle~II detector}\label{sec:detector}

The Belle~II detector~\cite{Abe:2010gxa} has a cylindrical geometry and includes a two-layer silicon-pixel detector~(PXD) surrounded by a four-layer double-sided silicon-strip detector~(SVD)~\cite{Belle-IISVD:2022upf} and a 56-layer central drift chamber~(CDC). These detectors reconstruct tracks of charged particles. The PXD and SVD are collectively referred to as the VXD, which is used to locate the decay vertex of B mesons. Only one sixth of the second layer of the PXD was installed for the data analyzed here. The $z$ axis is along the bisector of the angle between the direction of the electron beam and the direction opposite to the positron beam. The $x$-axis is horizontal and points away from the center of the accelerator ring. The $y$-axis is oriented vertically upward. With respect to the $z$-axis, $\phi$ is the azimuthal angle and $\theta$ is the polar angle. Surrounding the CDC, which also provides energy-loss measurements, is a time-of-propagation counter~(TOP)~\cite{Kotchetkov:2018qzw} in the central region and an aerogel-based ring-imaging Cherenkov counter~(ARICH) in the forward region. These detectors provide charged-particle identification. Surrounding the TOP and ARICH is the ECL, which is based on CsI(Tl) crystals and primarily provides energy and timing measurements for photons and electrons. The ECL covers the polar angle region of $12.4\degree<\theta<155.1\degree$, except for two gaps $\sim 1\degree$ wide between the barrel and endcap regions. Outside of the ECL is a superconducting solenoid magnet. Its flux return is instrumented with resistive-plate chambers and plastic scintillator modules to detect muons and $K^0_L$ mesons. The solenoid magnet provides a 1.5~T magnetic field that is parallel to the $z$ axis.

The Belle~II trigger system is designed to determine whether to retain or discard specific events. The L1 trigger operates as a hardware-based logic system with {a maximum output rate of 30~kHz and} a fixed latency of 4.4~${\rm \mu}$s~\cite{L1Trigger, L1Trigger2}. The HLT is a software-based system that reconstructs events and filters them to a maximum output rate of 10~kHz~\cite{hlt}. For each trigger bit in the HLT or L1, a prescale factor $R$ applies, recording only a fraction $1/R$ of the events that would have passed the trigger requirements. These prescale factors are taken into account in the analysis. 

\section{data sample and Monte Carlo simulation}\label{sec:data_mc}
The Run~1 data sample used in this paper was collected from March 2019 to June 2022 with the L1 trigger and HLT applied. Three categories of datasets were collected, corresponding to different
collision energies in the c.m.\ frame ($\sqrt{s}$), referred to as ``$\Upsilon(4S)$'' (10.580~GeV), ``off-$\Upsilon(4S)$'' ({10.517}~GeV), and ``$\Upsilon(5S)$ scan'' (10.657, 10.706, 10.751, and 10.810~GeV), where the values in parentheses are the c.m.\ energies for each dataset type. While we only present the details of the luminosity measurement for the $\Upsilon(4S)$ dataset in this paper, we apply the same measurement technique to all other datasets.

To determine the selection efficiencies, three signal Monte Carlo (MC) samples are generated at the c.m.\ energy corresponding to the peak of the $\Upsilon(4S)$ resonance. The BABAYAGA@NLO generator~\cite{Bhabha-gen, digamma-gen} is used to produce events equivalent to 36~fb$^{-1}$ for the Bhabha process and 719~fb$^{-1}$ for the digamma process. The KKMC generator~\cite{Jadach:1999vf} is used to generate 1439~fb$^{-1}$ of dimuon events. Taking into account the beam energy variation of approximately 3.1~MeV for the LER and 4.4~MeV for the HER, an energy spread of 5~MeV for $\sqrt{s}$ is applied in the generation of Bhabha, digamma, and dimuon events. The scattering polar angles of the final-state particles are set to be within the range of 10.0\degree~to 170.0\degree~in the lab frame, fully covering the detector acceptance. The cross-sections in this angular window obtained from the generators are $\sigma_{ee}$ = ({295.38 $\pm$ 0.04 $\pm$ 0.42})~nb, $\sigma_{\gamma \gamma}$ = ({5.0686 $\pm$ 0.0005 $\pm$ 0.0071})~nb and $\sigma_{\mu \mu}$ = ({1.1472 $\pm$ 0.0001 $\pm$ 0.0051})~nb, where the first uncertainties are statistical and the second are systematic. To evaluate the impact of background contributions, nine types of background processes are generated: 200~fb$^{-1}$ equivalent samples of $B^+B^-$ and $B^0{\bar B^0}$ events utilizing the EvtGen generator~\cite{EvtGen}; 200~fb$^{-1}$ equivalent samples of $u{\bar u}$, $d{\bar d}$, $c{\bar c}$, and $s{\bar s}$ quark continuum events generated via a combination of KKMC~\cite{Jadach:1999vf} and PYTHIA~\cite{PYTHIA}; a 1~ab$^{-1}$ equivalent sample of $ \tau ^+ \tau ^-$ events generated with KKMC~\cite{KKMC-taupair1,KKMC-taupair2}; and 100~fb$^{-1}$ equivalent samples of $e^+e^-e^+e^-$ and $e^+e^- \mu ^+ \mu ^-$ events produced using AAFH~\cite{AAFH-eeXX}. After MC event generation, the interactions of final-state particles with detector material are simulated with GEANT4~\cite{Agostinelli:2002hh}, and the subsequent subdetector response and L1 and HLT triggers are simulated with the Belle II Analysis Software Framework (basf2)~\cite{Kuhr:2018lps}. Time-dependent MC simulations, incorporating calibrated beam background, {interaction point (IP)} position, and other time-dependent quantities aligned with actual data, are employed to determine the luminosity. Time-dependent MC simulations are used to assess some systematic uncertainties as outlined in Section~\ref{sec:sys_uncer}.

\section{Event selection}\label{sec:evt_sel}

\begin{figure*}
  \centering
  \begin{tabular}{c c c}
  \includegraphics[width=0.33\linewidth]{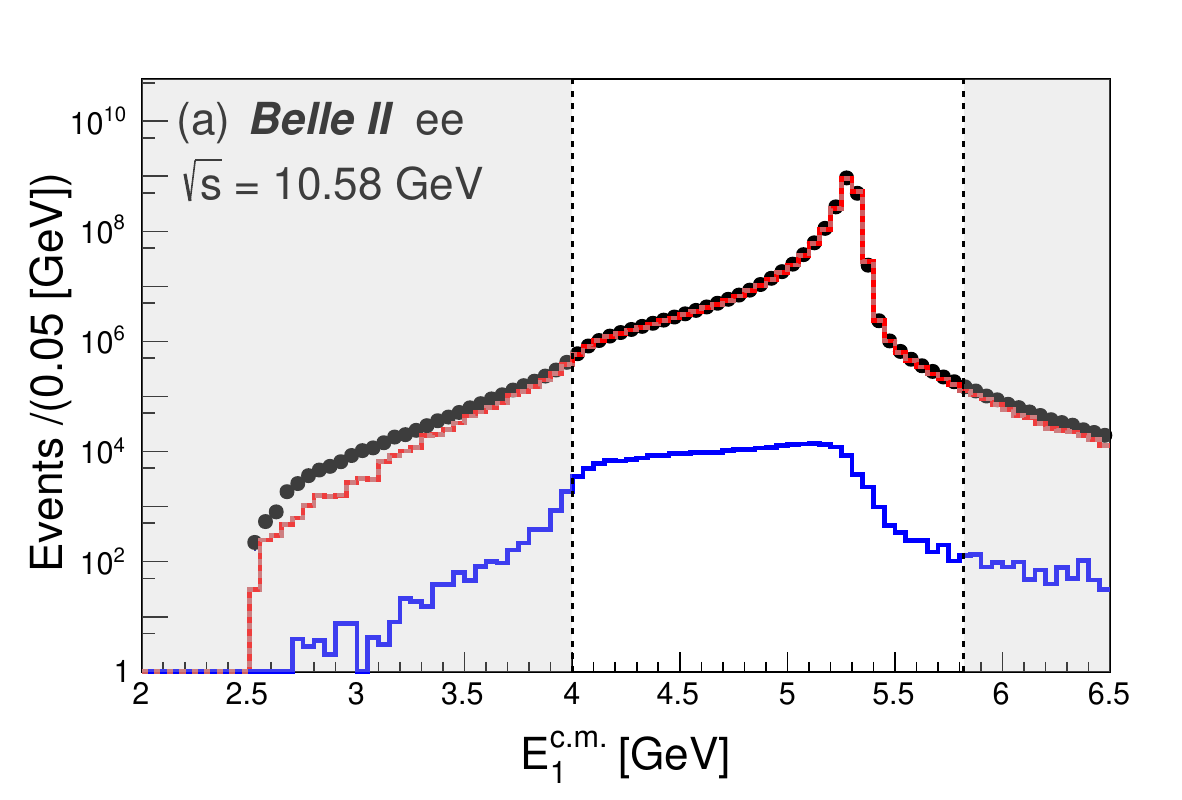}&
  \includegraphics[width=0.33\linewidth]{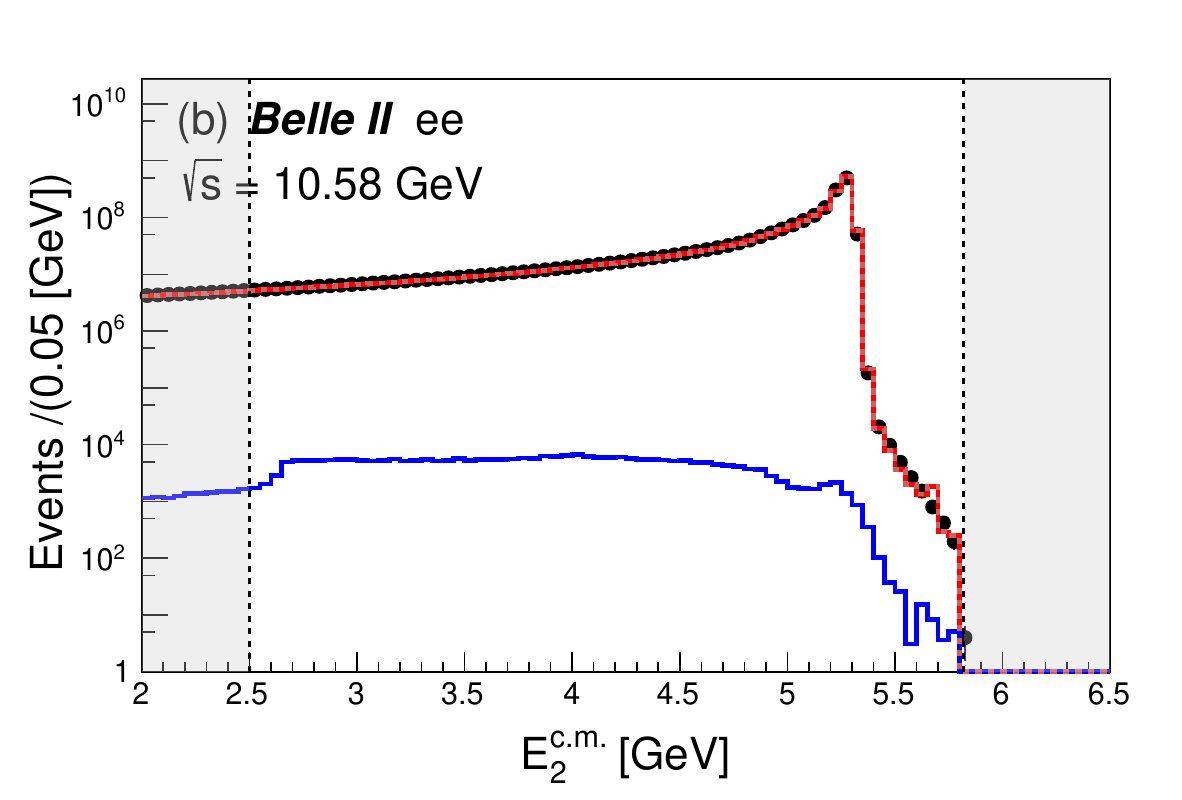}&
  \includegraphics[width=0.33\linewidth]{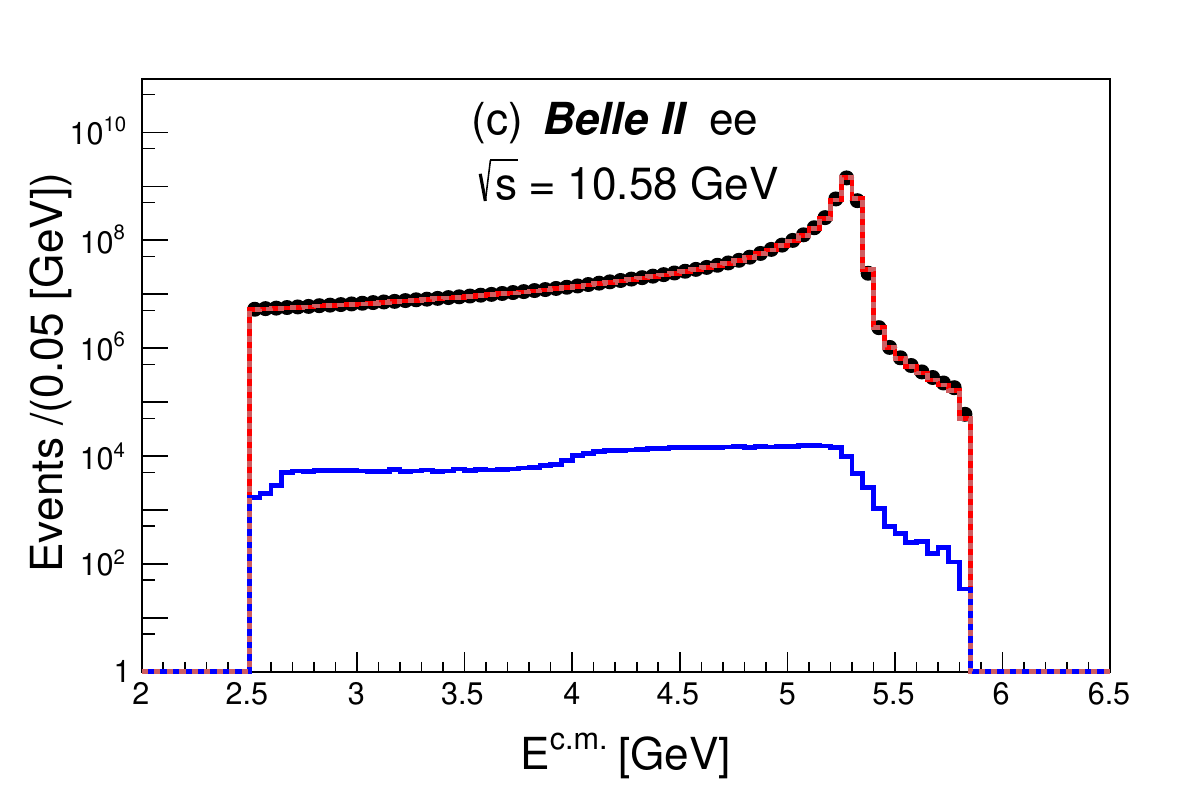}\\
  \includegraphics[width=0.33\linewidth]{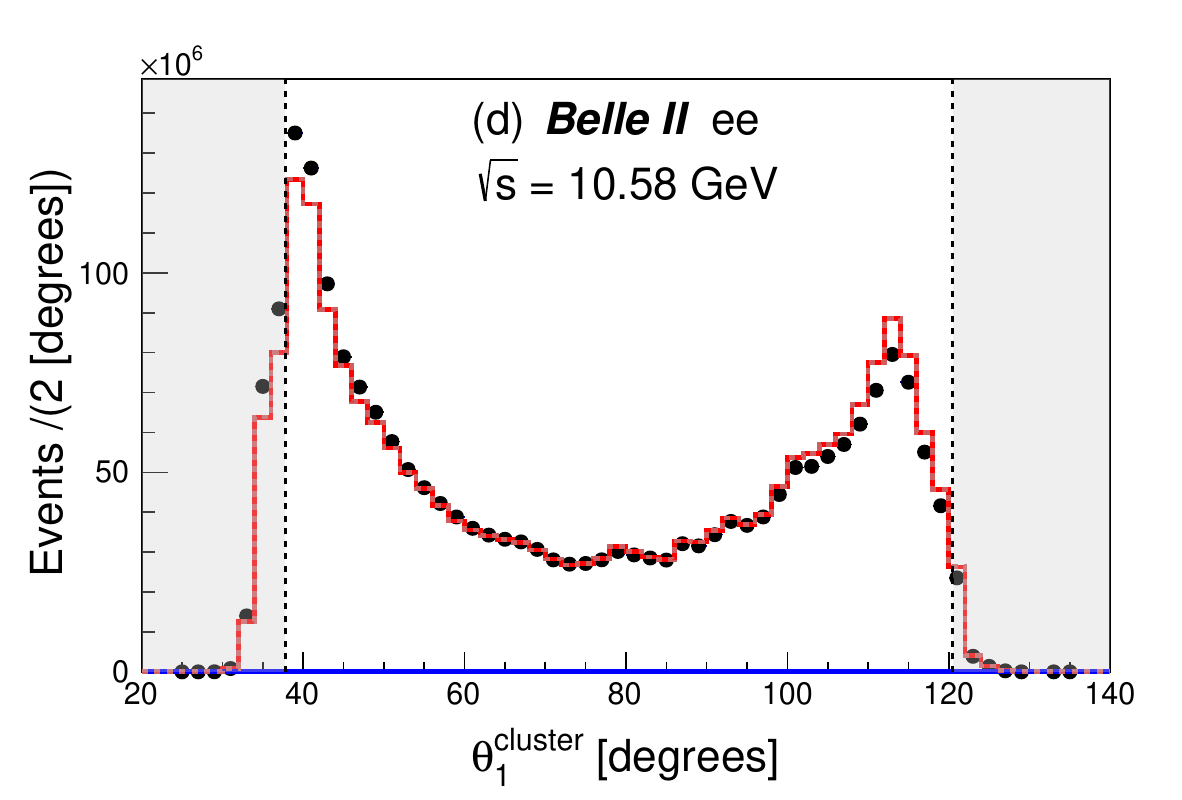}&
  \includegraphics[width=0.33\linewidth]{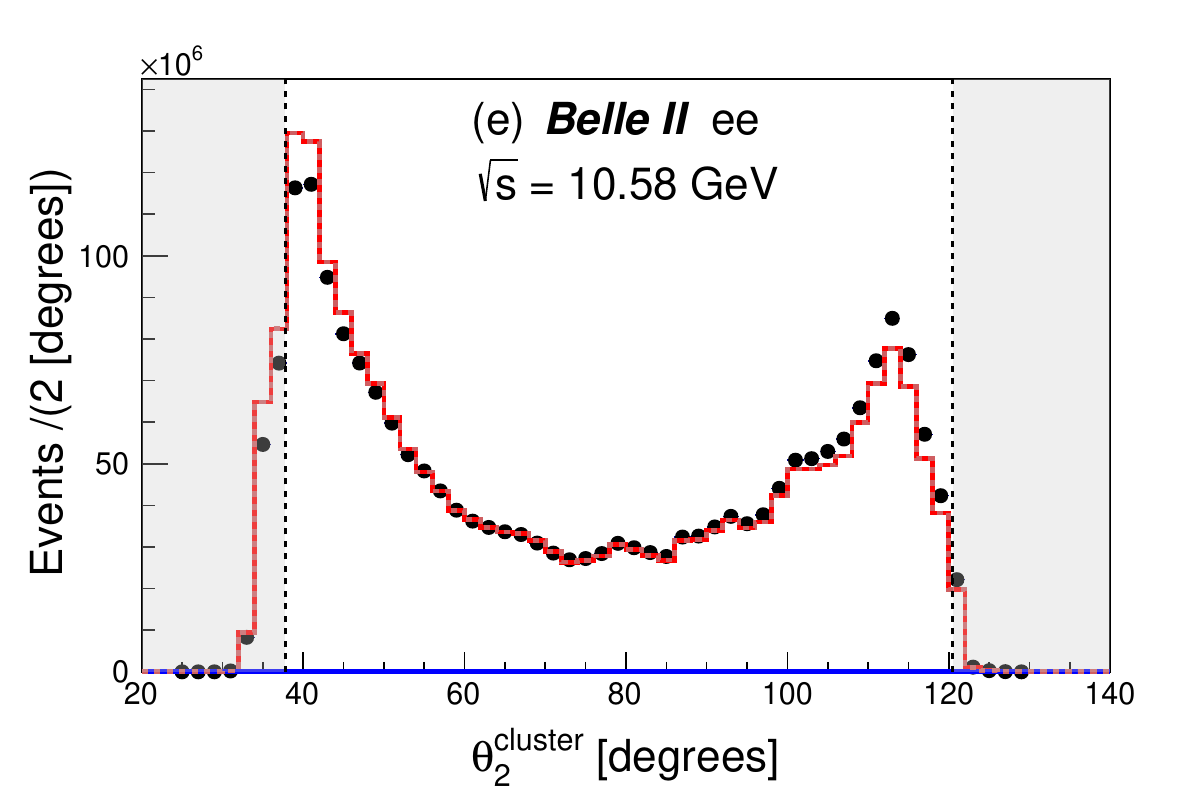}&
  \includegraphics[width=0.33\linewidth]{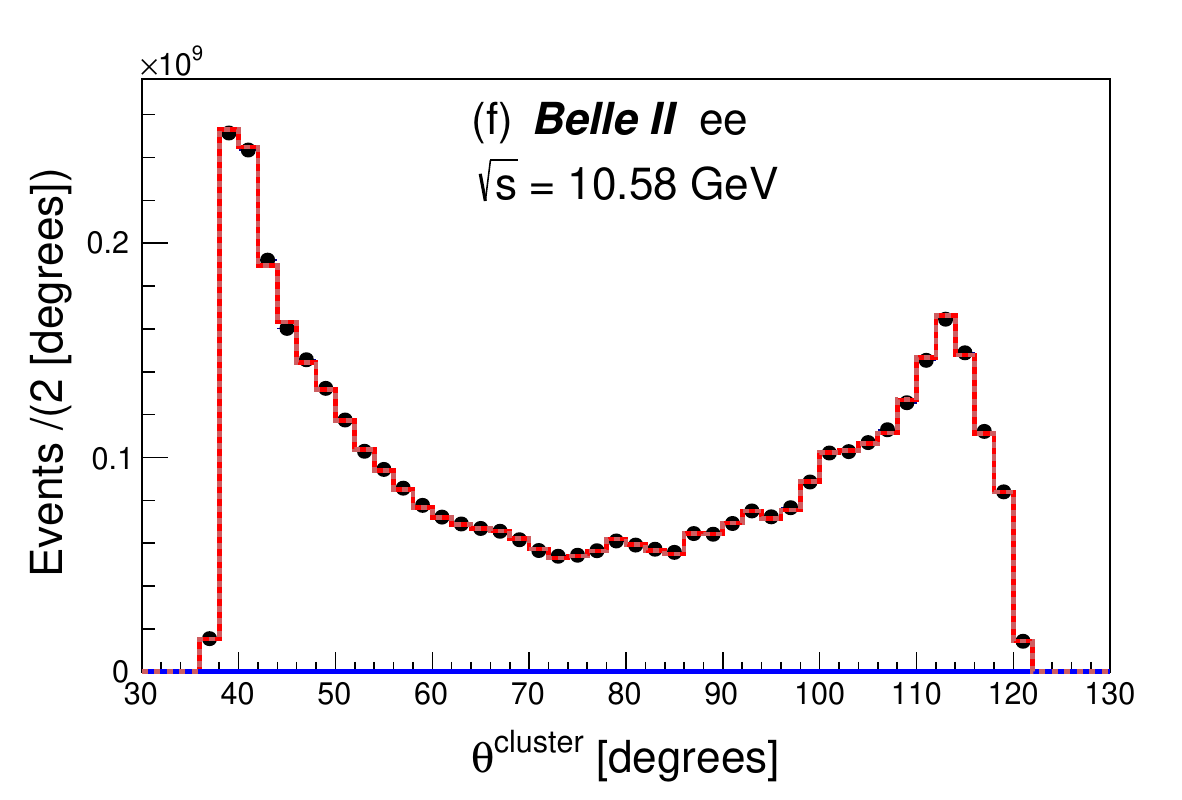}\\
  \includegraphics[width=0.33\linewidth]{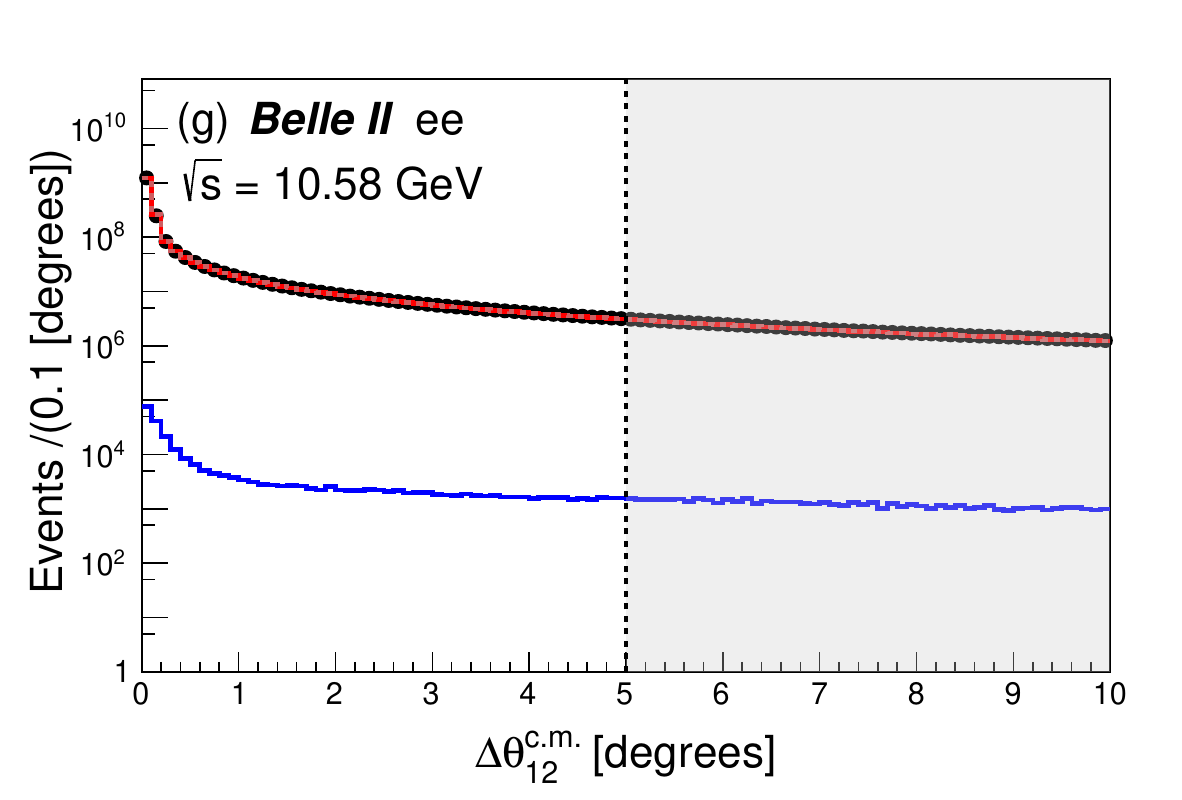}&
  \includegraphics[width=0.33\linewidth]{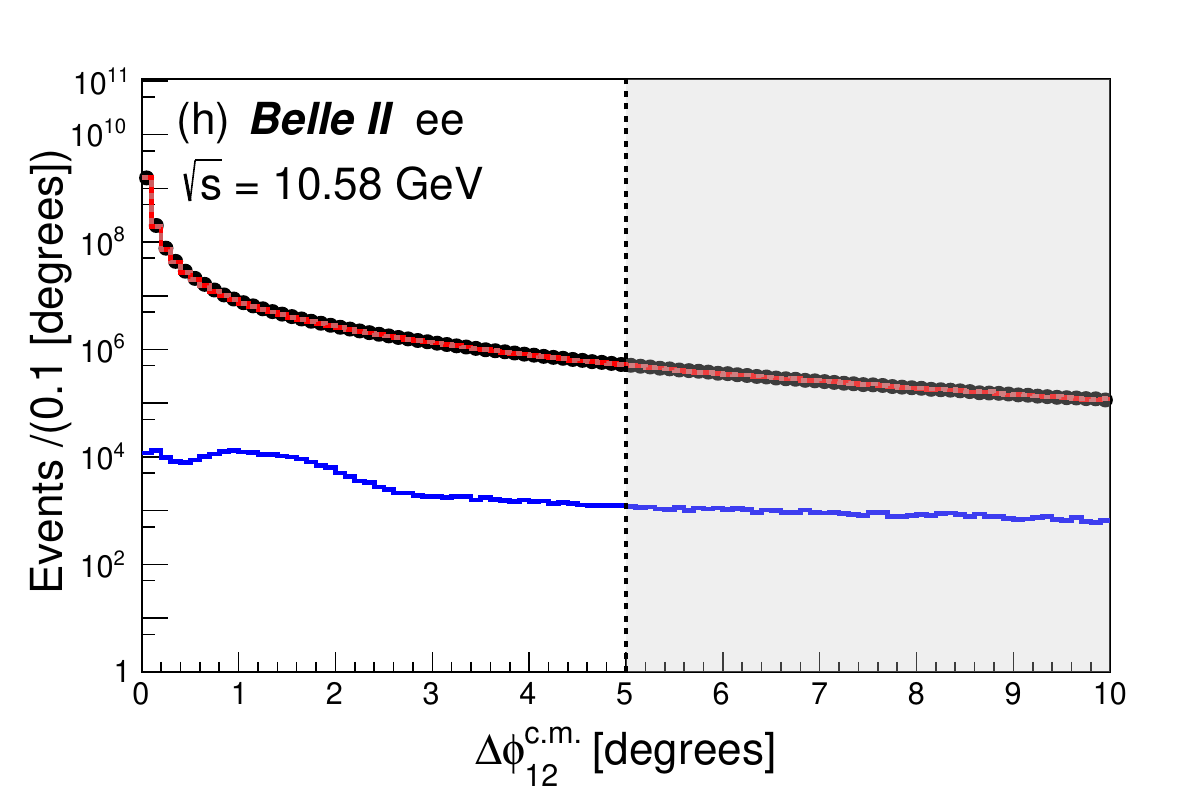}&
  \includegraphics[width=0.33\linewidth]{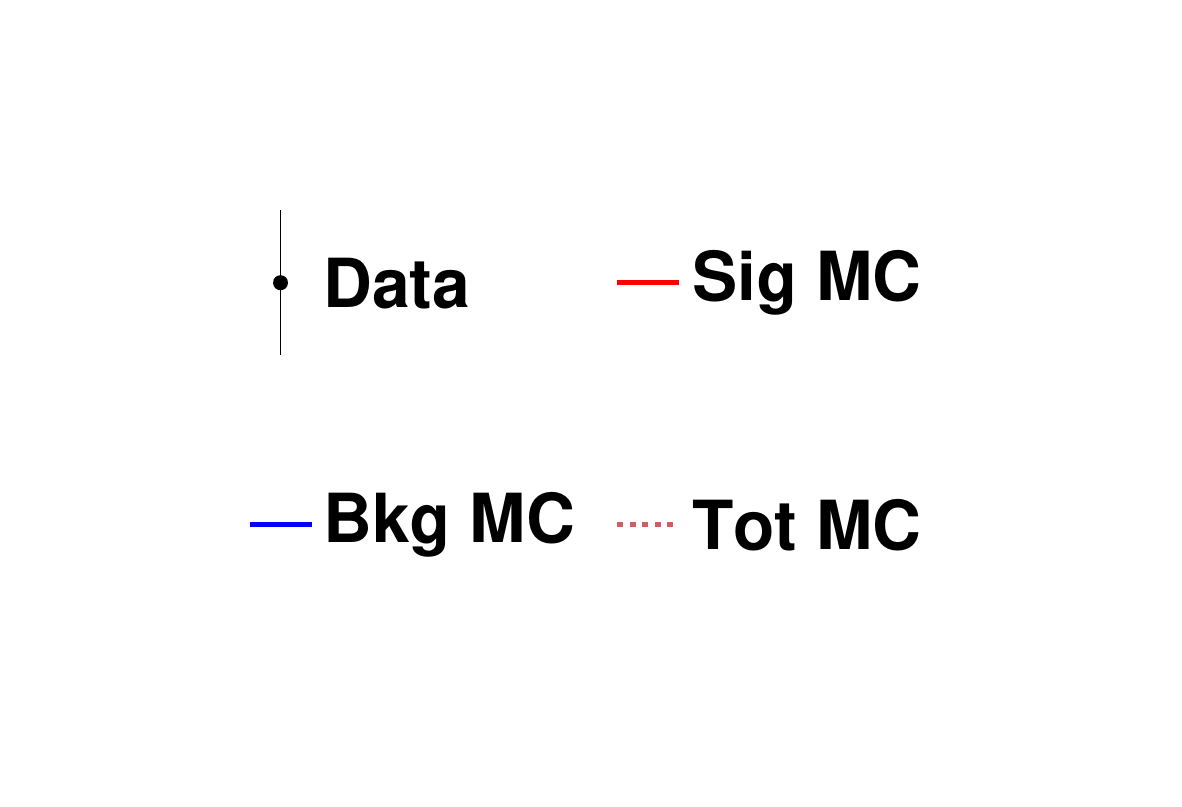}\\
  \end{tabular}
  \caption{Kinematic distributions of Bhabha signal candidates in data and MC. The first row contains distributions of the c.m.\ energy of the higher-energy final-state particle (a), the lower-energy final-state particle (b), and each of the two final-state particles (c). The second row represents distributions of polar angle of ECL clusters of the higher-energy final-state particle (d), the lower-energy particle (e), and each of the two final-state particles (f) in the lab frame. Subfigure (g) shows the distribution of the polar opening angle ($\Delta\theta_{12}^{\rm c.m.}=|\theta_{\rm 1}^{\rm c.m.}+\theta_{\rm 2}^{\rm c.m.} -180\degree|$), while subfigure (h) shows the distribution of the azimuthal opening angle ($\Delta\phi^{\rm c.m.}_{12} = ||\phi^{\rm c.m.}_{\rm 1}-\phi^{\rm c.m.}_{\rm 2}|-180\degree|$). Each subfigure shows one parameter from the event selection criteria where all other criteria have been applied except for that shown in the figure. All the criteria are applied in subfigures (c) and (f). ``Data'' in the legend signifies the collision data from the $\Upsilon(4S)$ dateset, while ``Sig MC'', ``Bkg MC'', and ``Tot MC'' correspond to the signal (MC Bhabha events), background (MC samples except for signal events), and total MC samples (the sum of both the signal and background samples), respectively. The error bars are invisible due to the large size of the sample. The dashed histogram for ``Tot MC" overlaps with the solid histogram of ``Sig MC", making the ``Tot MC" histogram invisible. The vertical black dashed lines represent the selection criteria for signal events. Events in the shaded regions are removed by the selection criteria.}\label{fig:dis_bha_4S}
\end{figure*}

\begin{figure*}
 \centering
 \begin{tabular}{c c c}
  \includegraphics[width=0.33\linewidth]{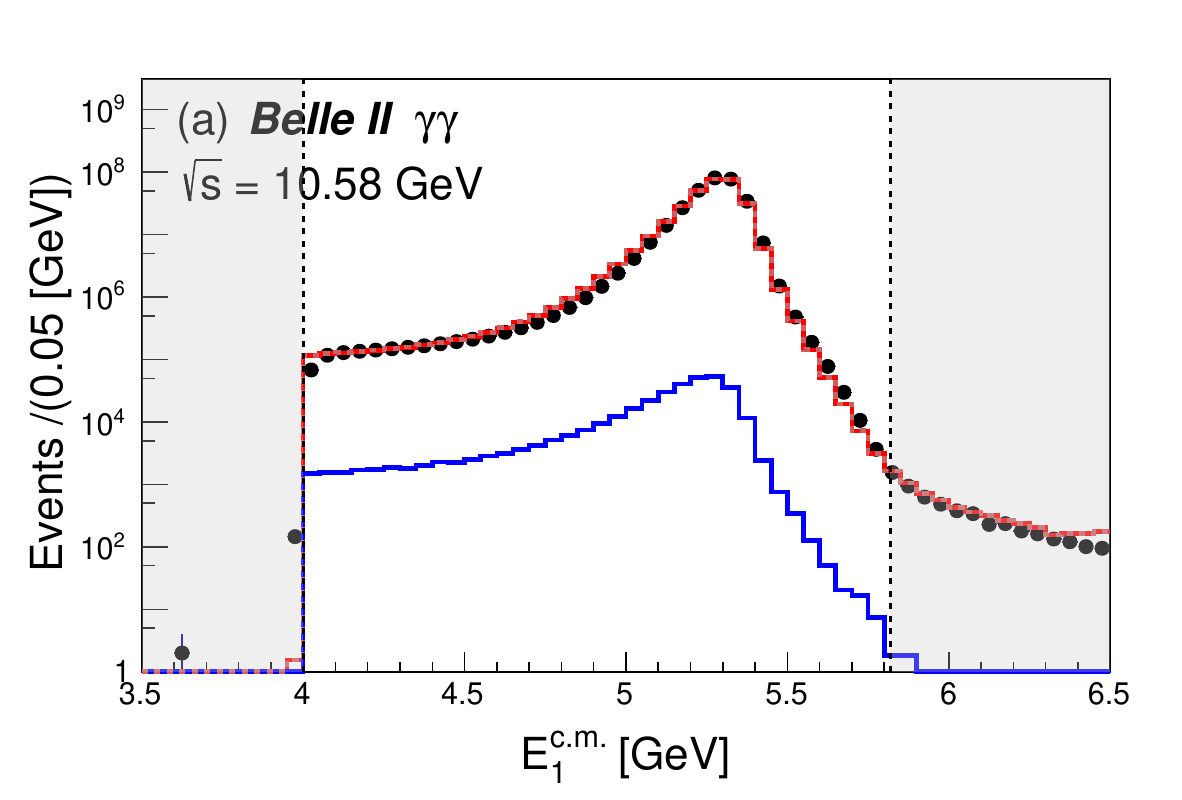}&
  \includegraphics[width=0.33\linewidth]{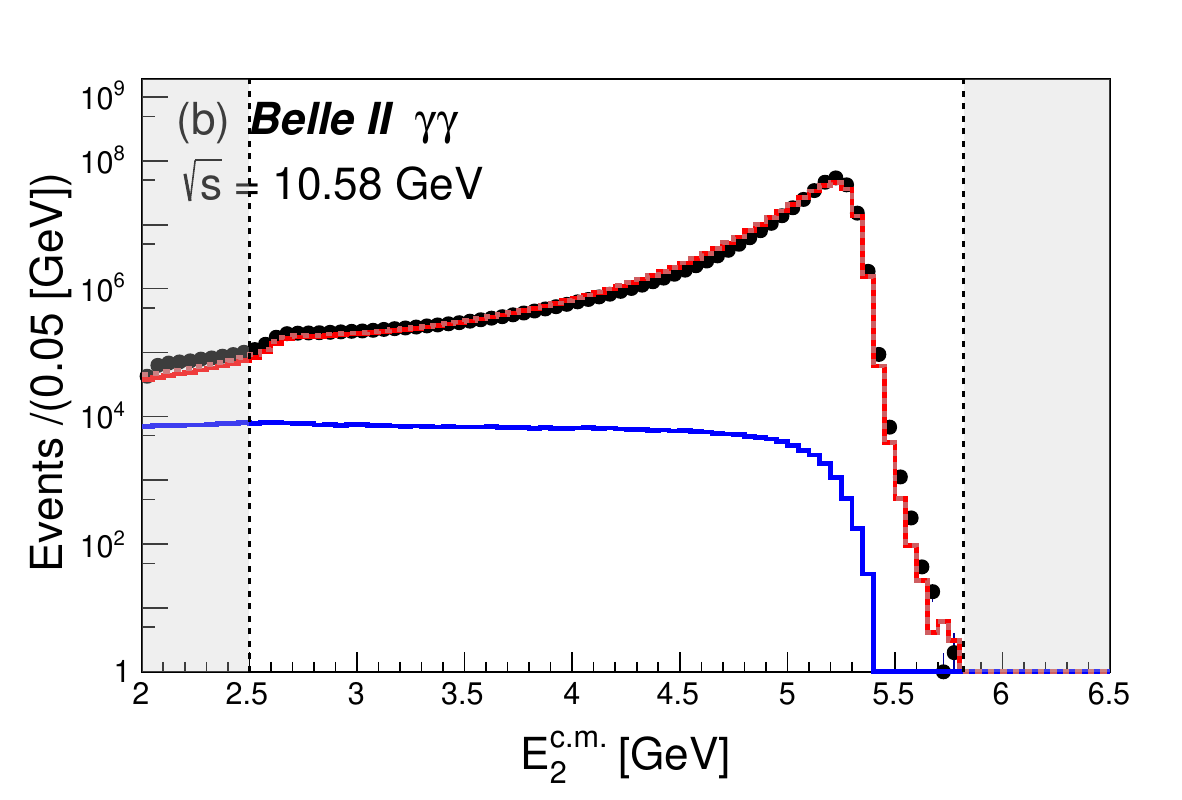}&
  \includegraphics[width=0.33\linewidth]{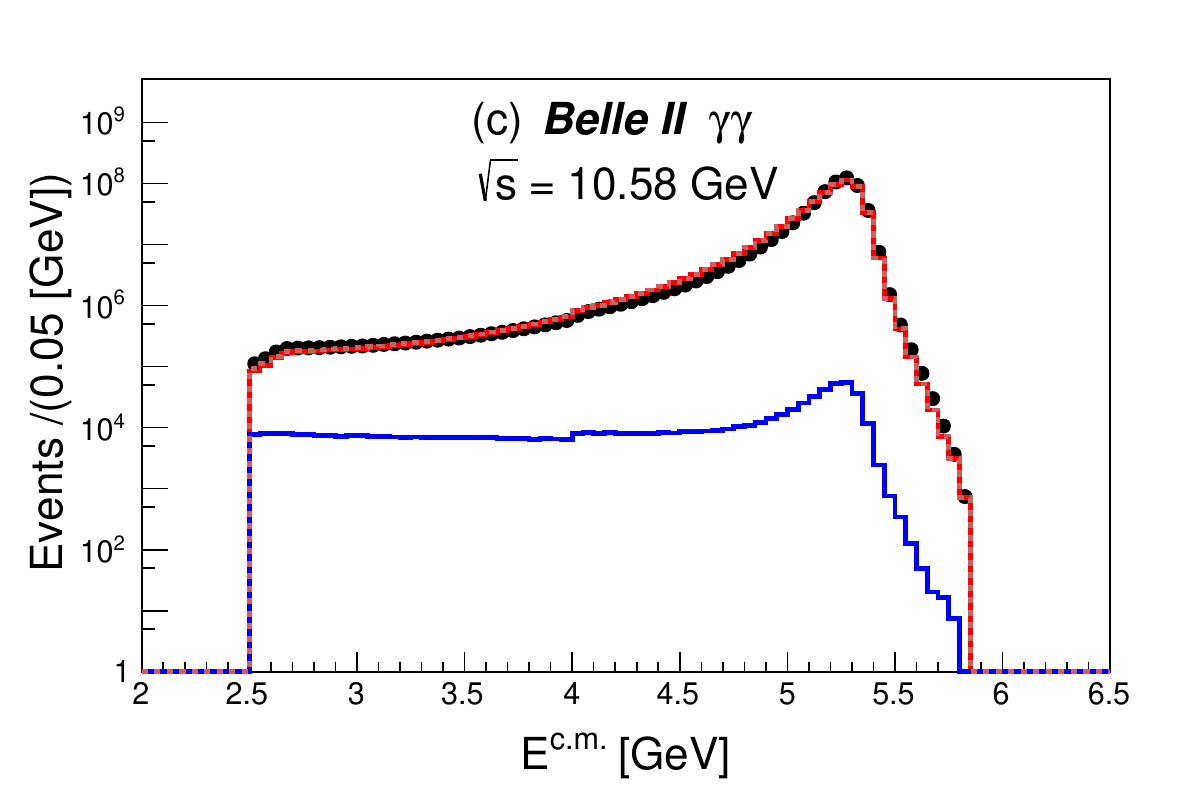}\\
  \includegraphics[width=0.33\linewidth]{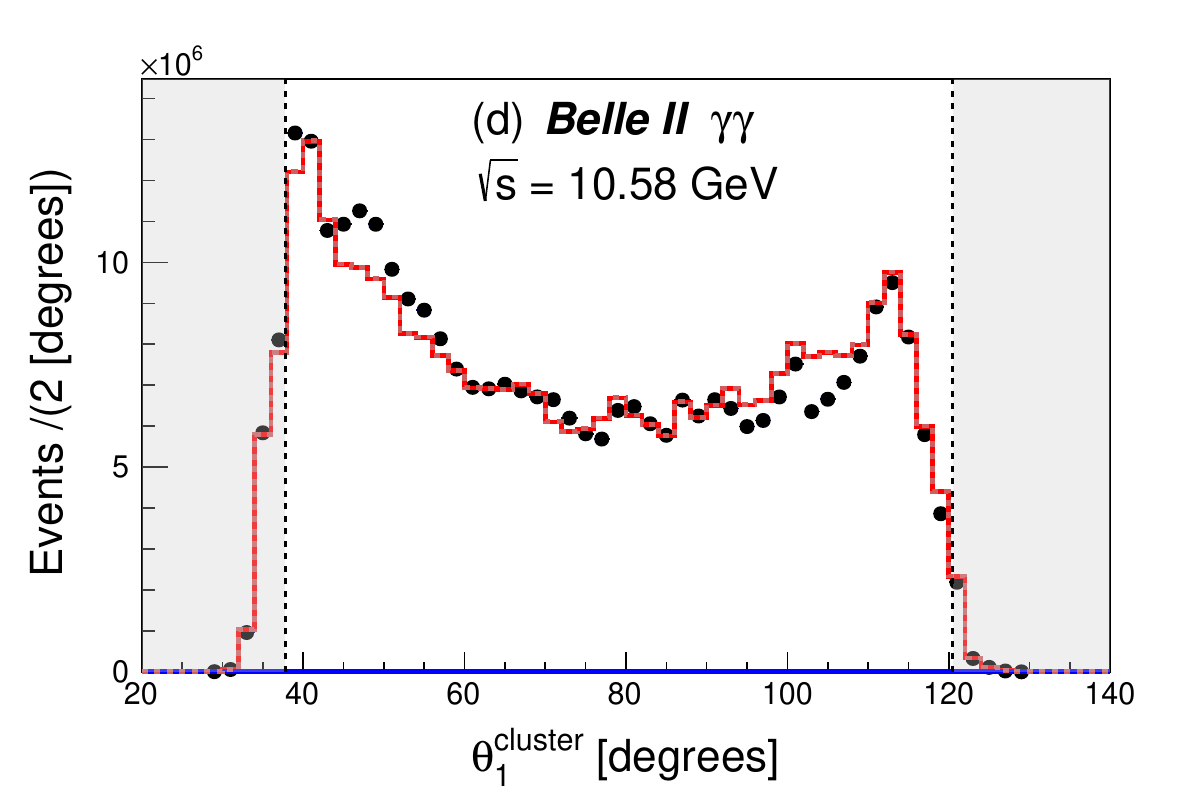}&
  \includegraphics[width=0.33\linewidth]{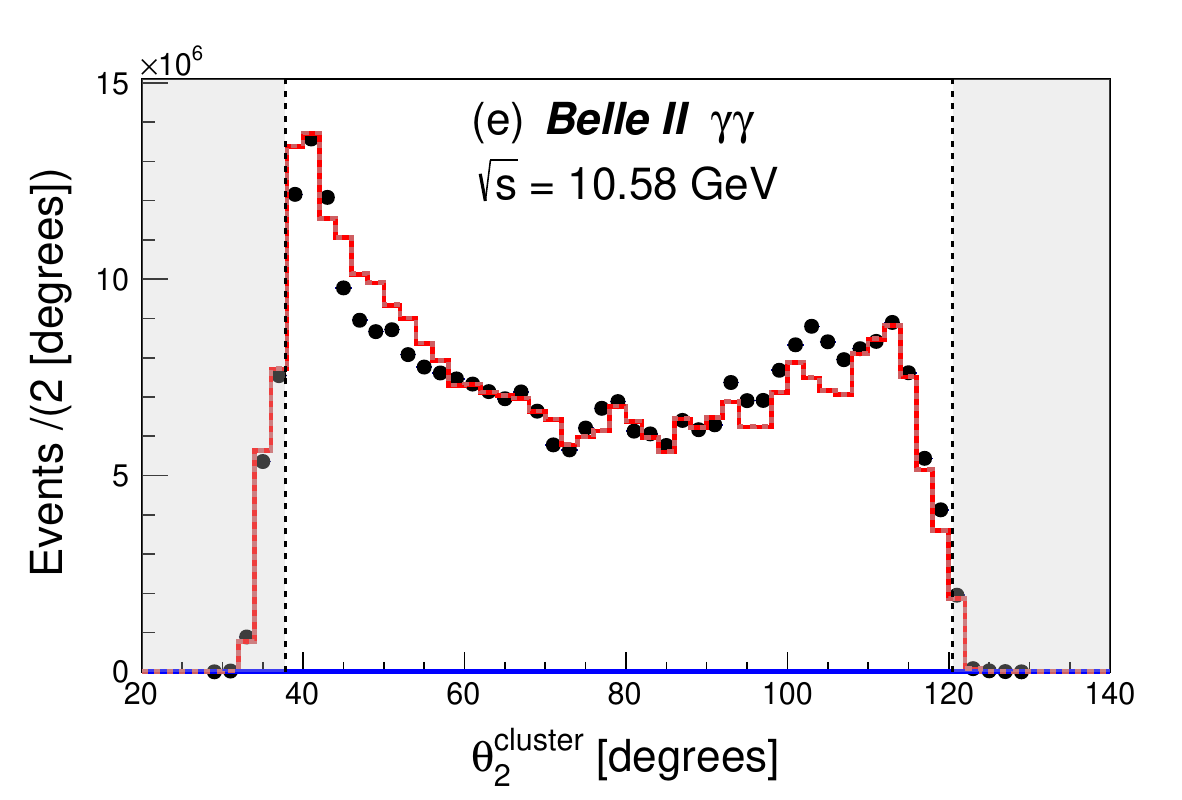}&
  \includegraphics[width=0.33\linewidth]{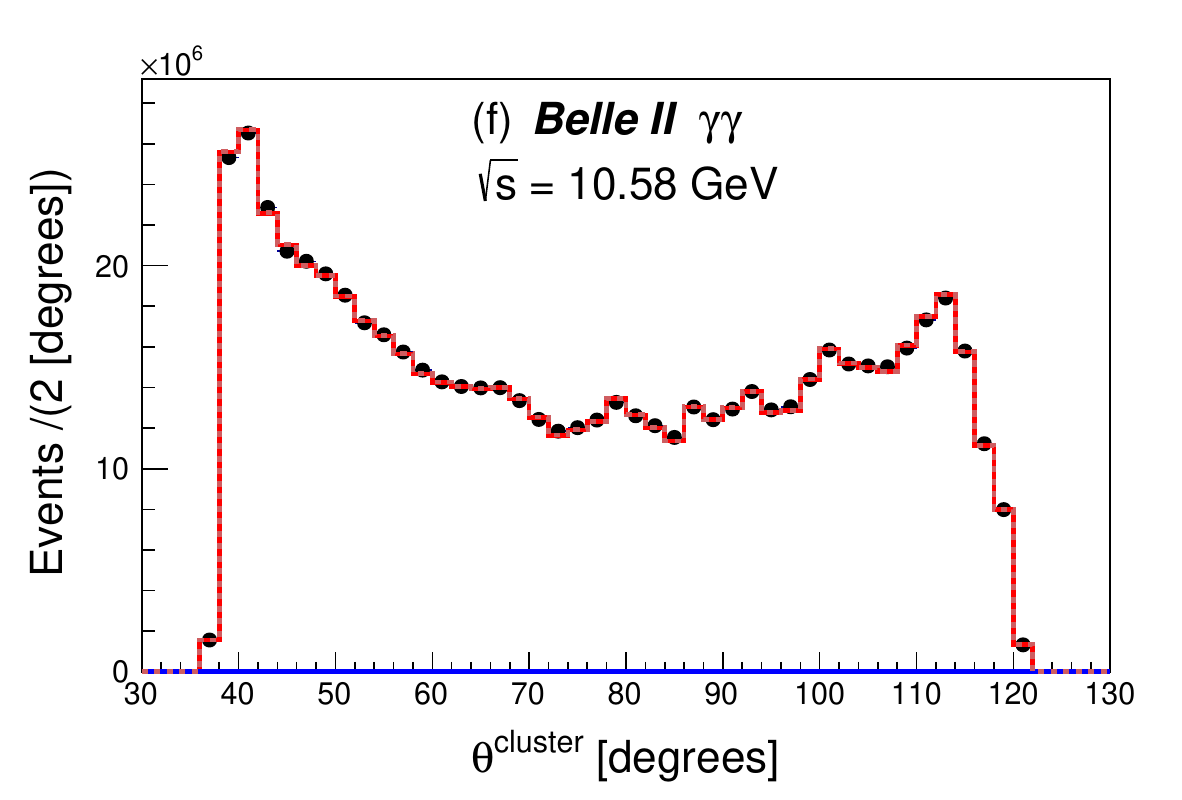}\\
  \includegraphics[width=0.33\linewidth]{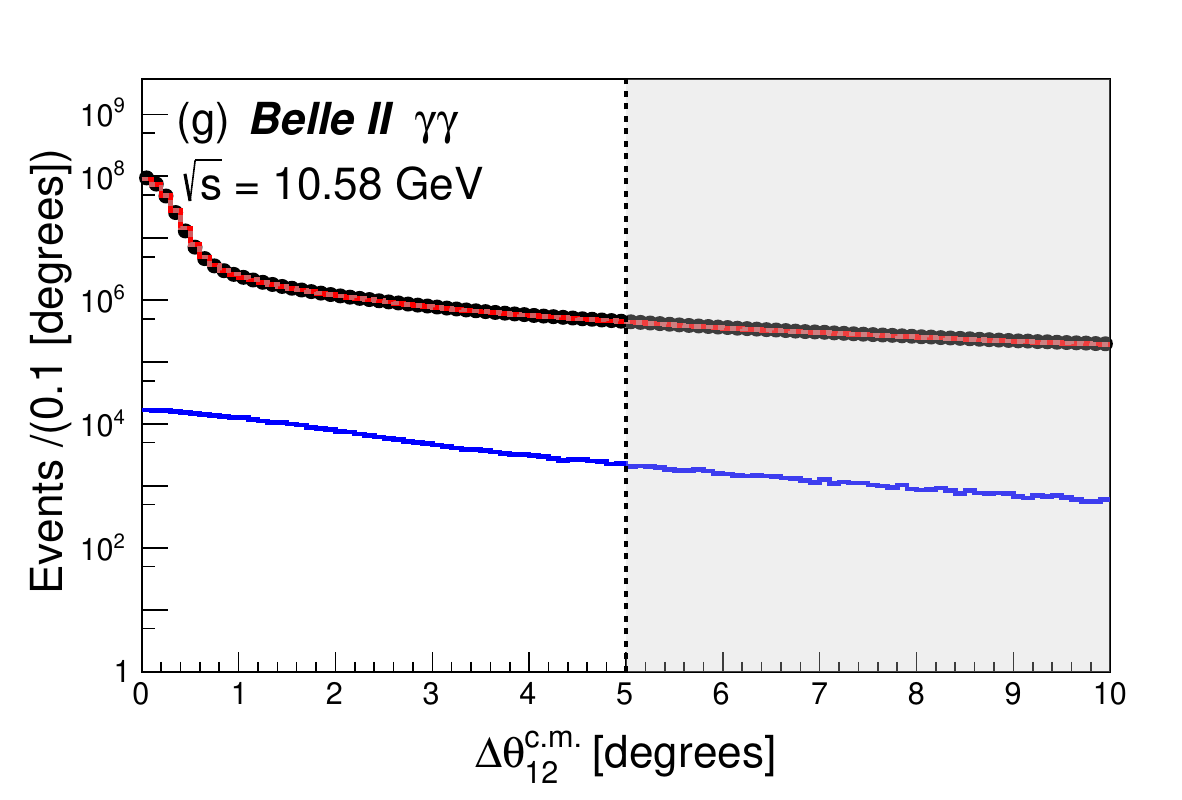}&
  \includegraphics[width=0.33\linewidth]{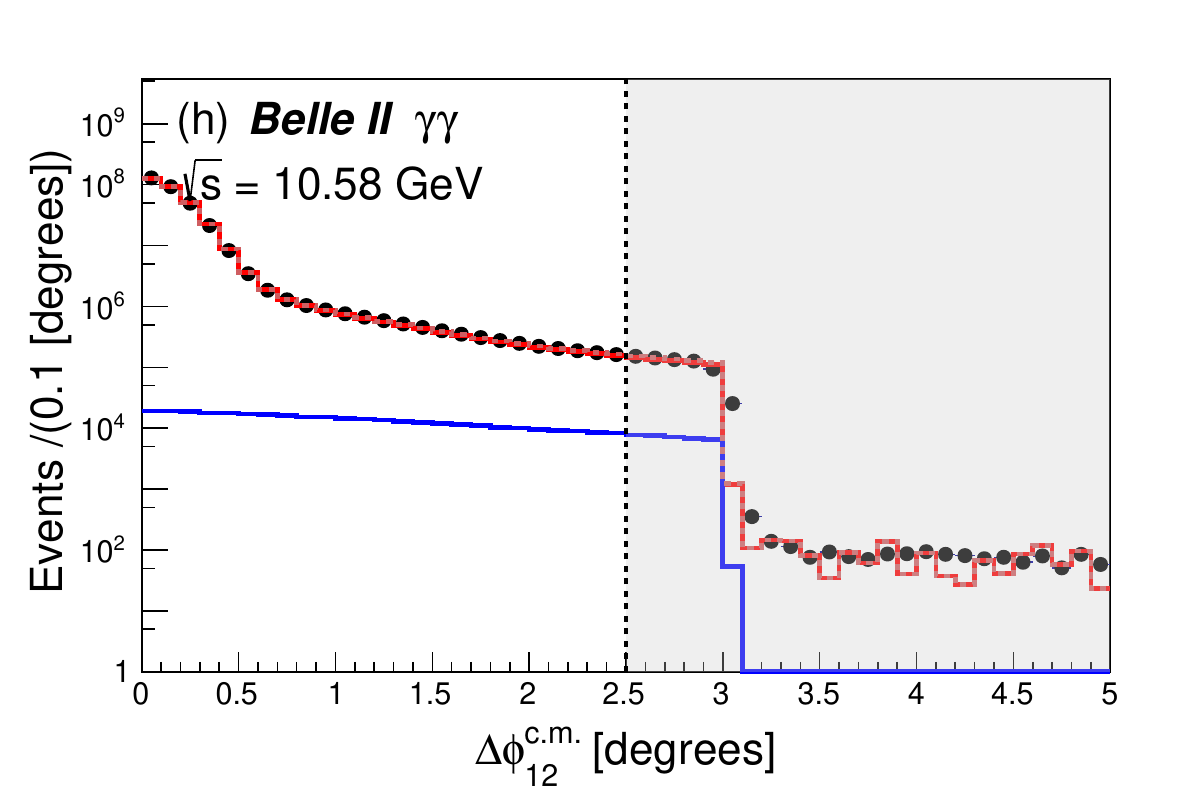}&
  \includegraphics[width=0.33\linewidth]{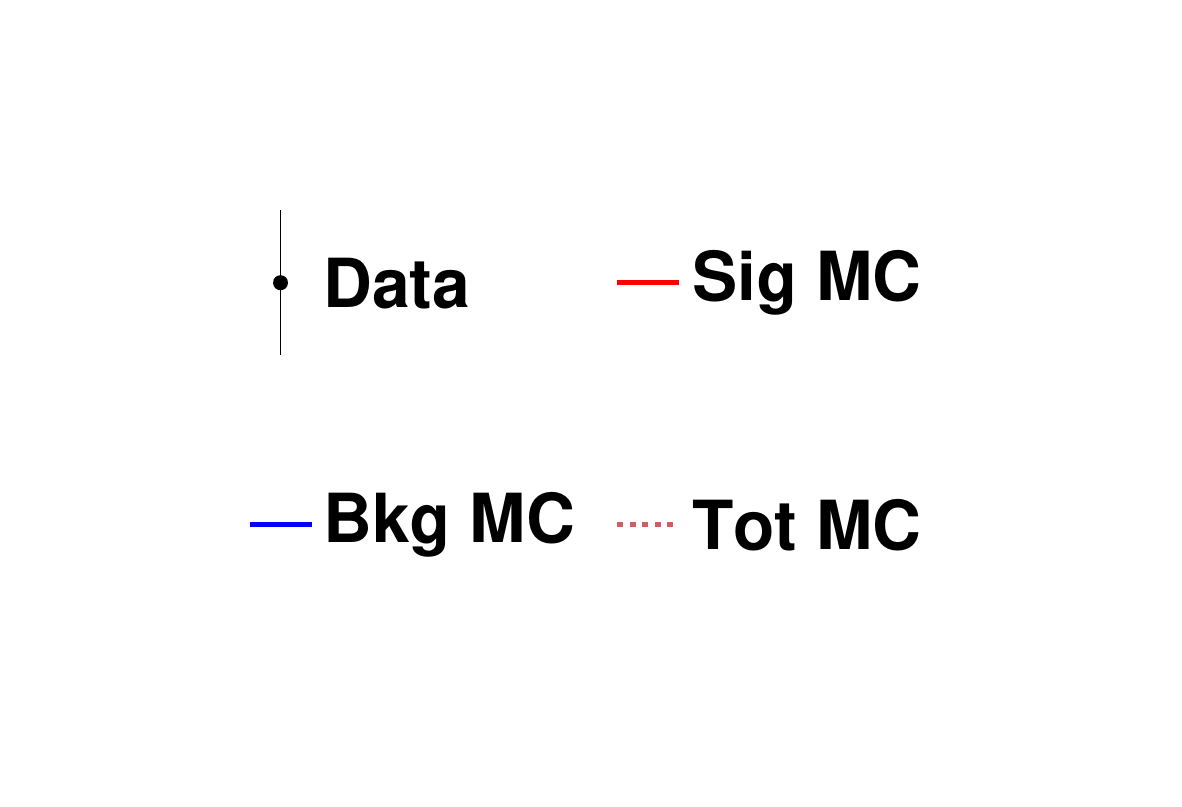}\\
  \end{tabular}
  \caption{Kinematic distributions of digamma-dominated signal candidates in data and MC. The convention in the figure is the same as in Fig.~\ref{fig:dis_bha_4S} except the signal processes are Bhabha and digamma events here. The edge at 3$^\circ$ in the $\Delta \phi_{12}^{\rm c.m.}$ distribution is due to the HLT requirements for digamma-dominated channel discussed in the text.}\label{fig:dis_dig_4S}
\end{figure*}

\begin{figure*}
    \centering
    \begin{tabular}{c c c}
  \multicolumn{3}{c}{\includegraphics[width=0.9\linewidth]{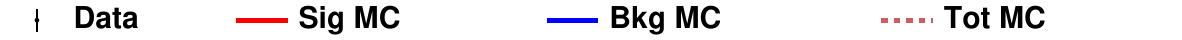}}\\
  \includegraphics[width=0.33\linewidth]{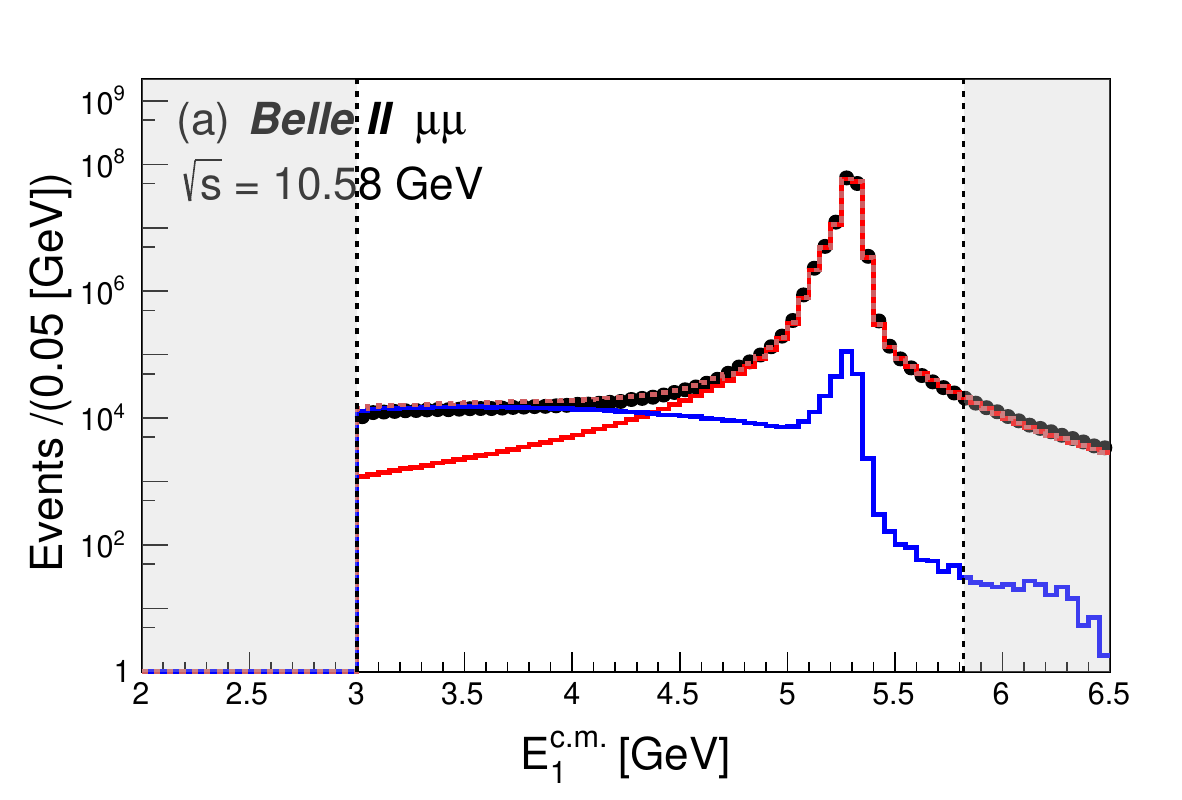}&
  \includegraphics[width=0.33\linewidth]{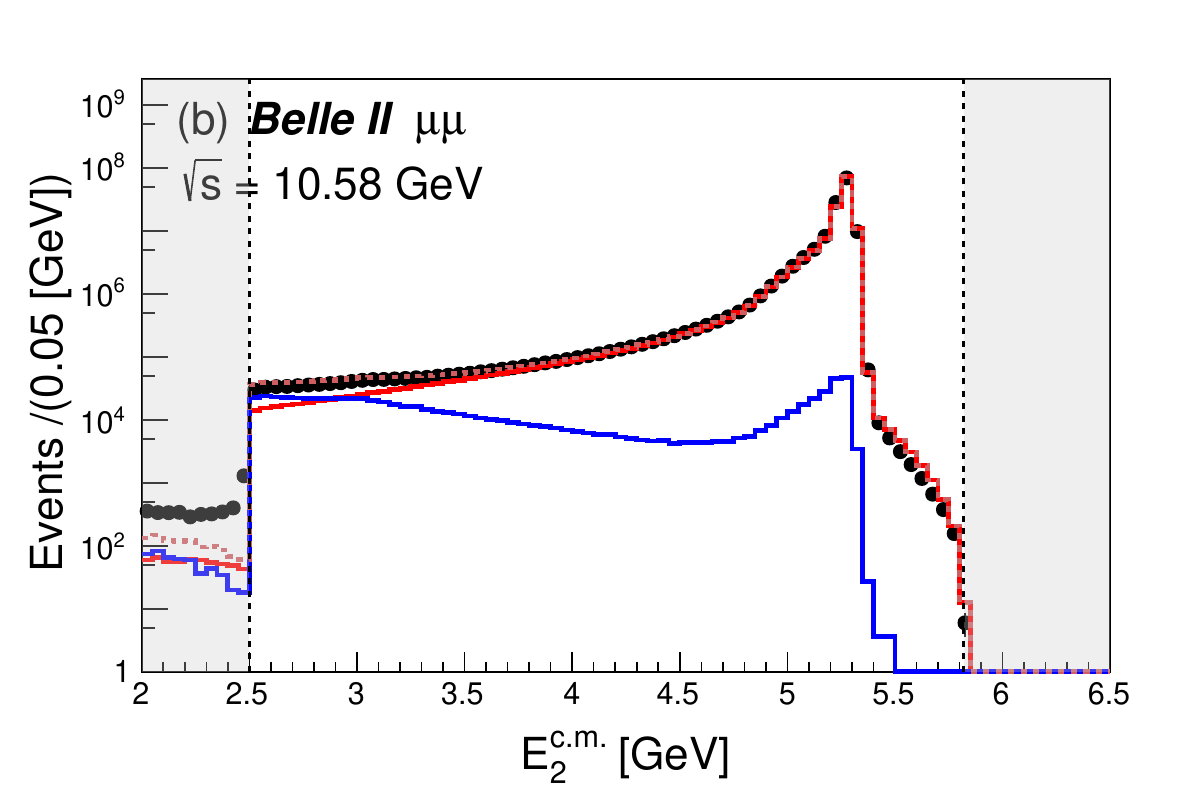}&
  \includegraphics[width=0.33\linewidth]{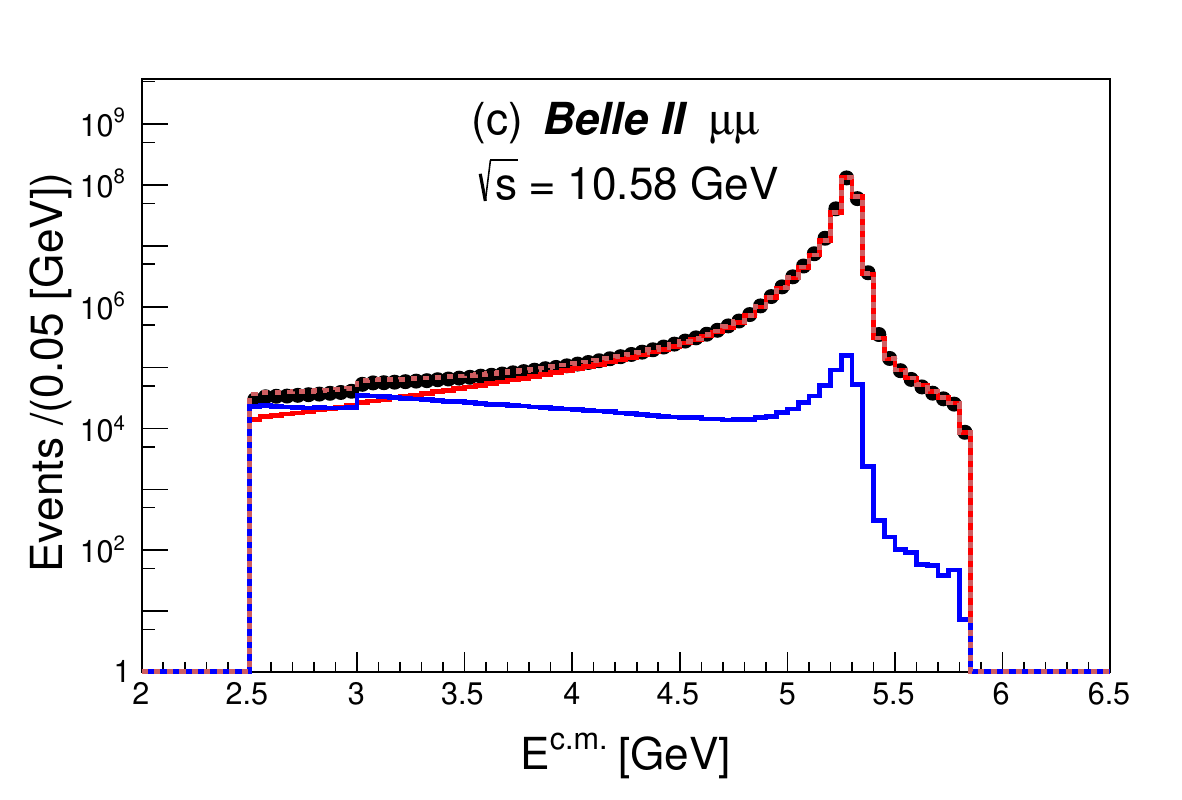}\\
  \includegraphics[width=0.33\linewidth]{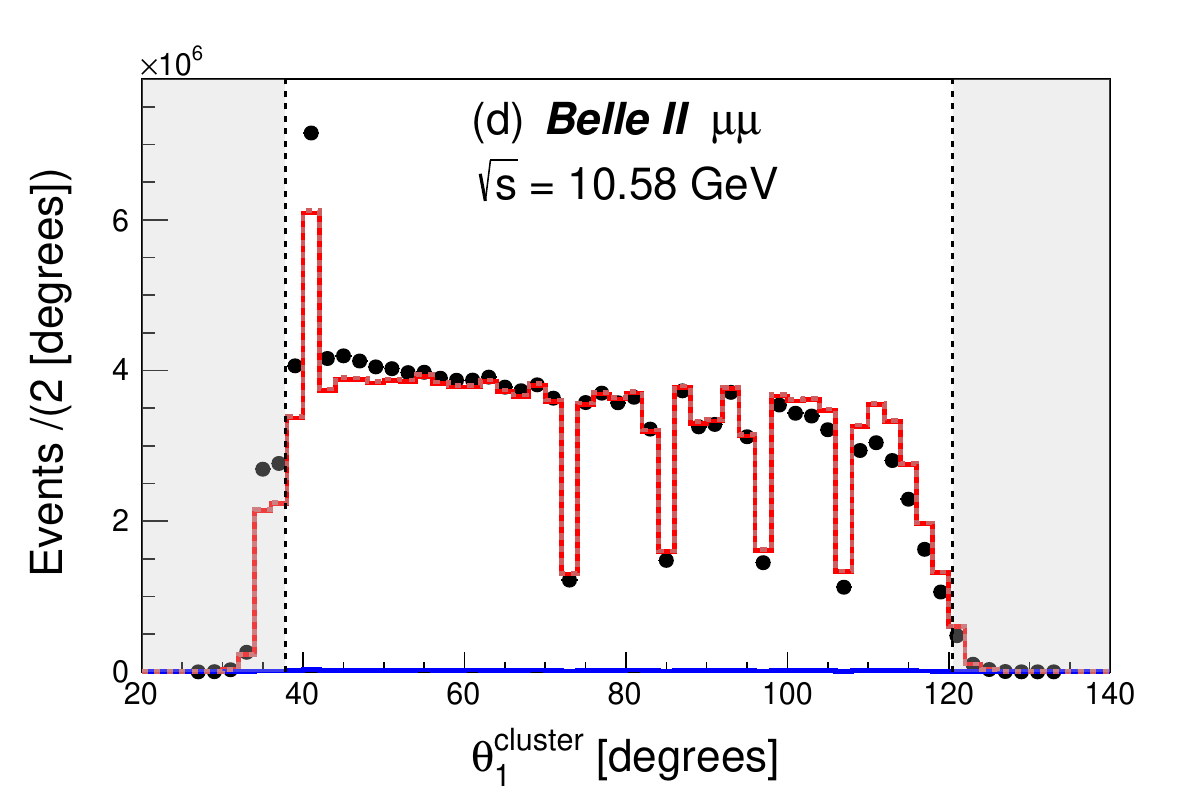}&
  \includegraphics[width=0.33\linewidth]{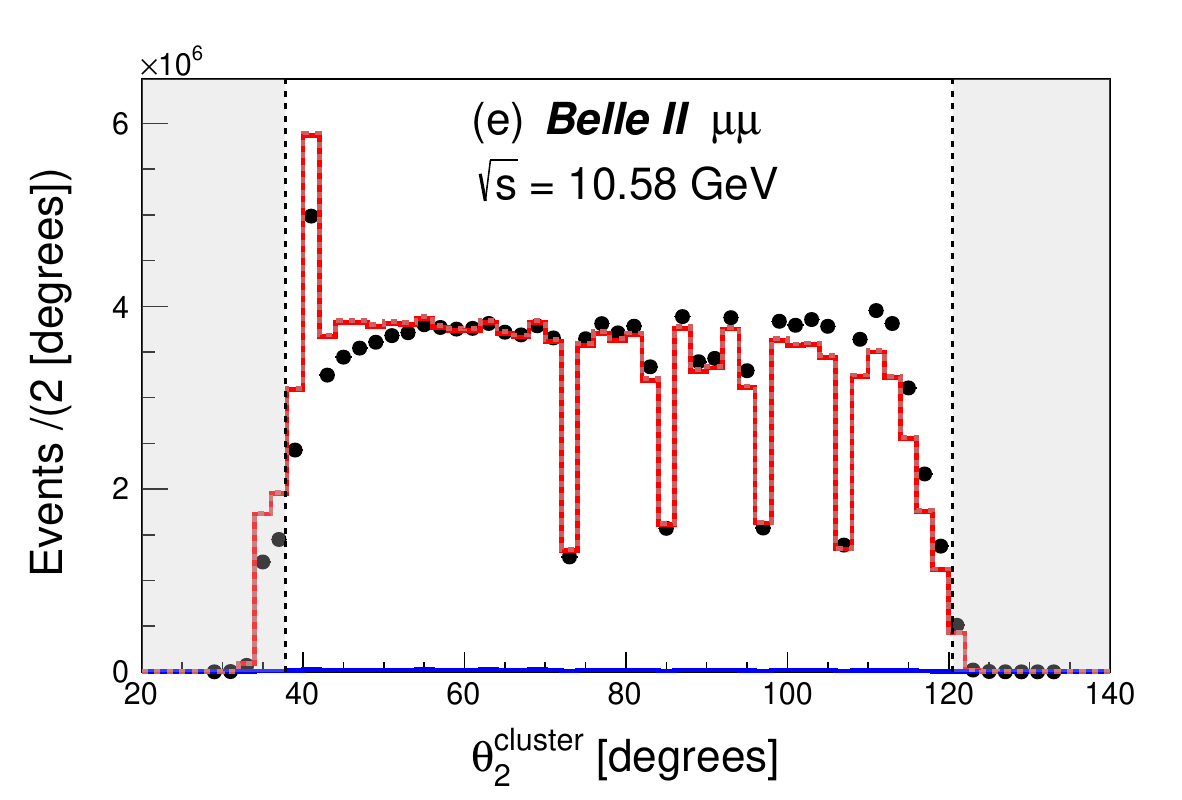}&
  \includegraphics[width=0.33\linewidth]{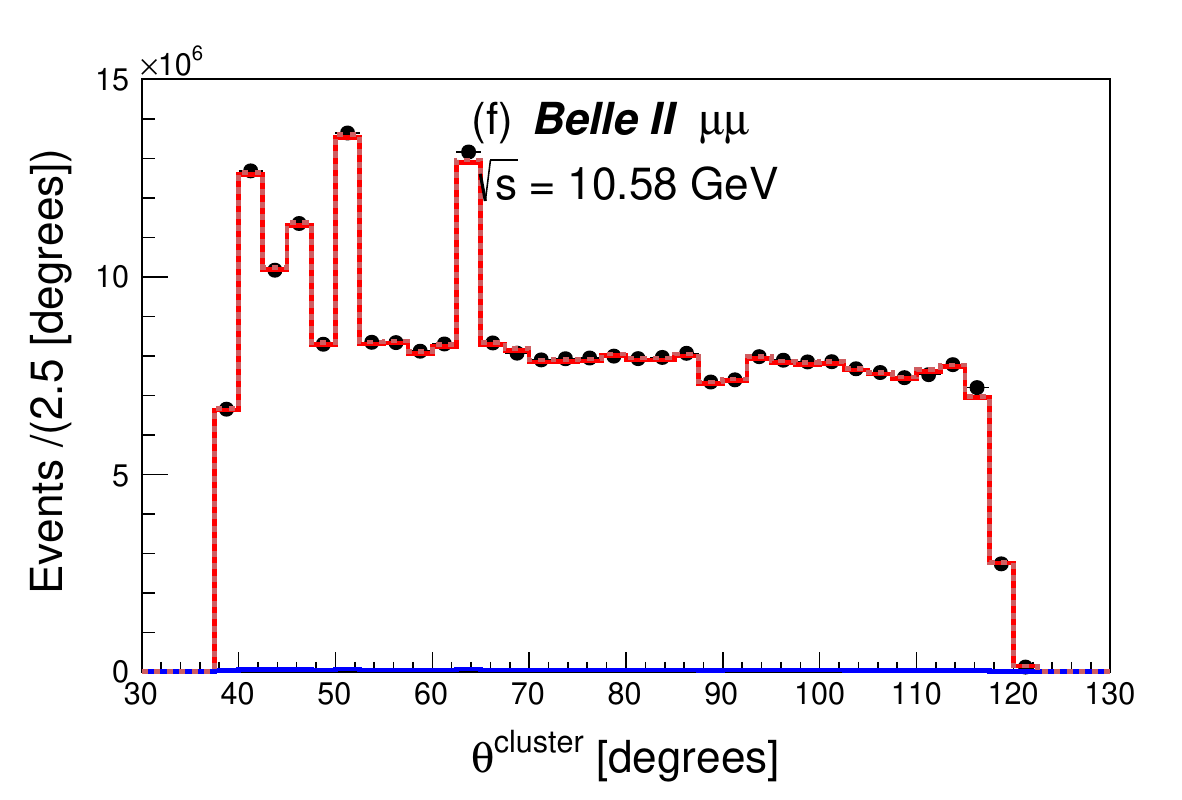}\\
  \includegraphics[width=0.33\linewidth]{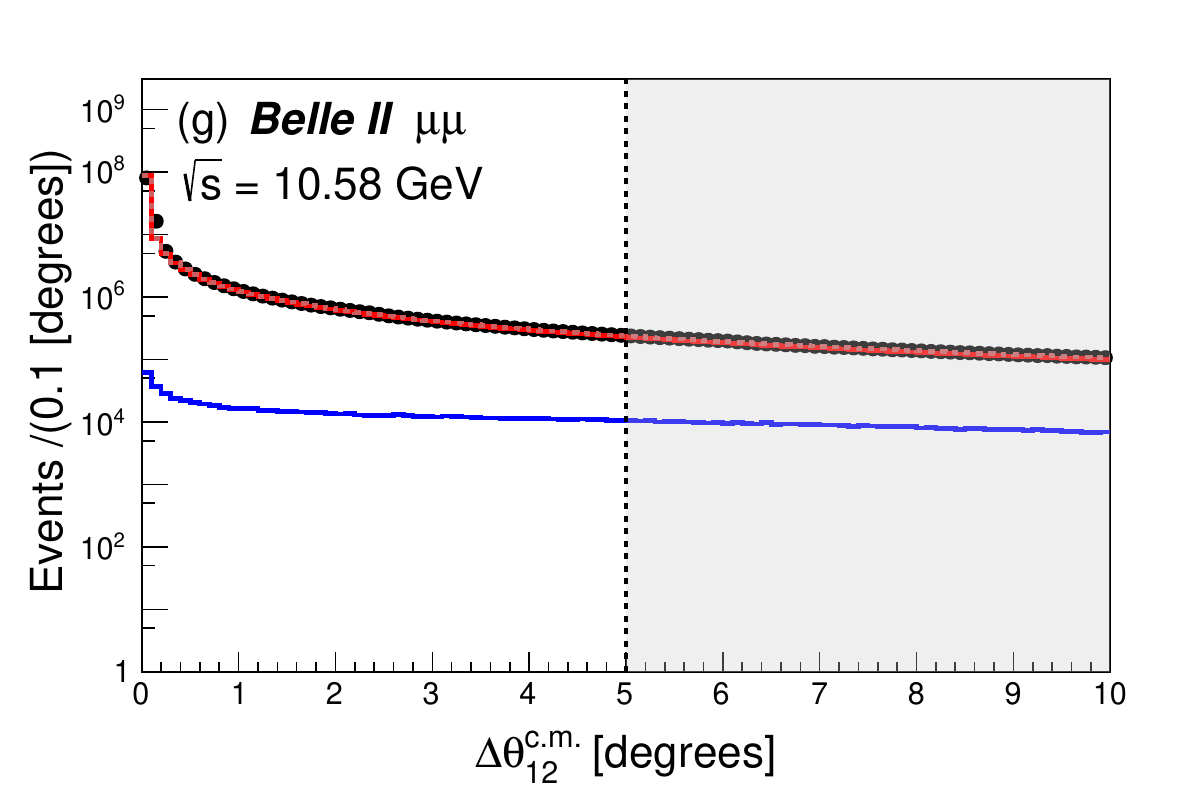}&
  \includegraphics[width=0.33\linewidth]{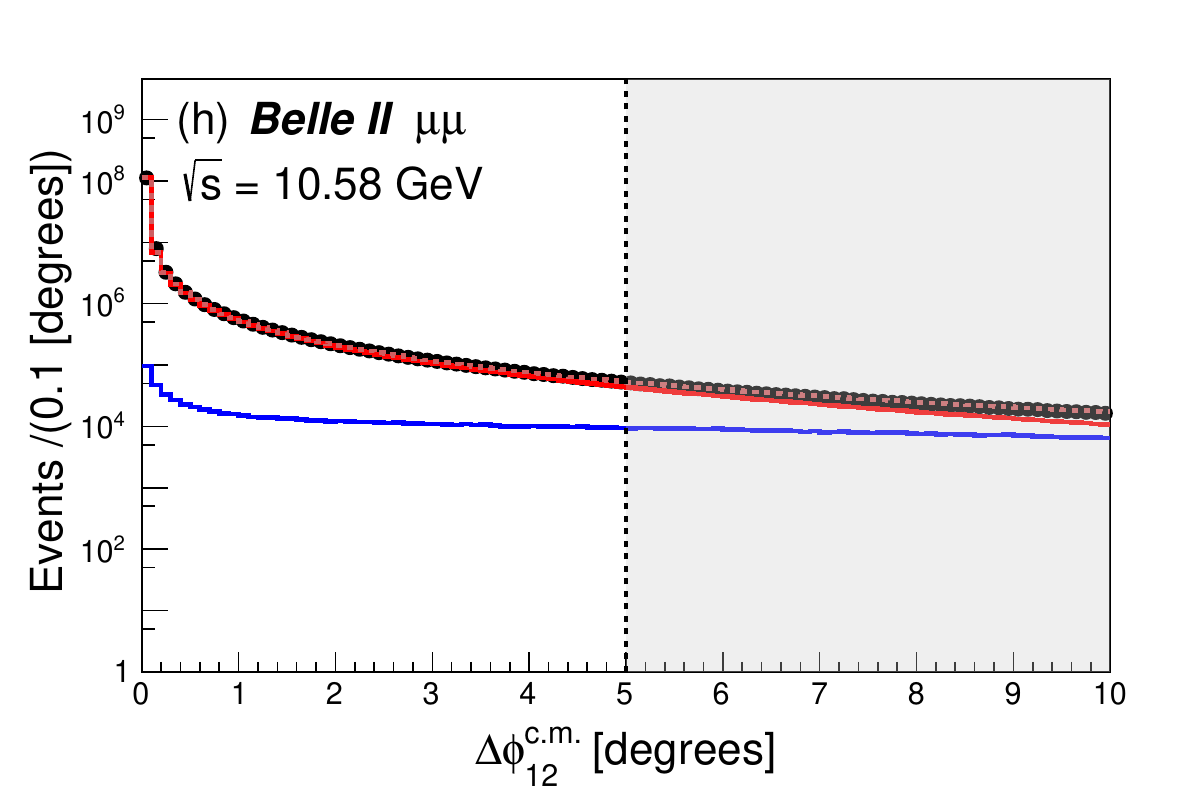}&
  \includegraphics[width=0.33\linewidth]{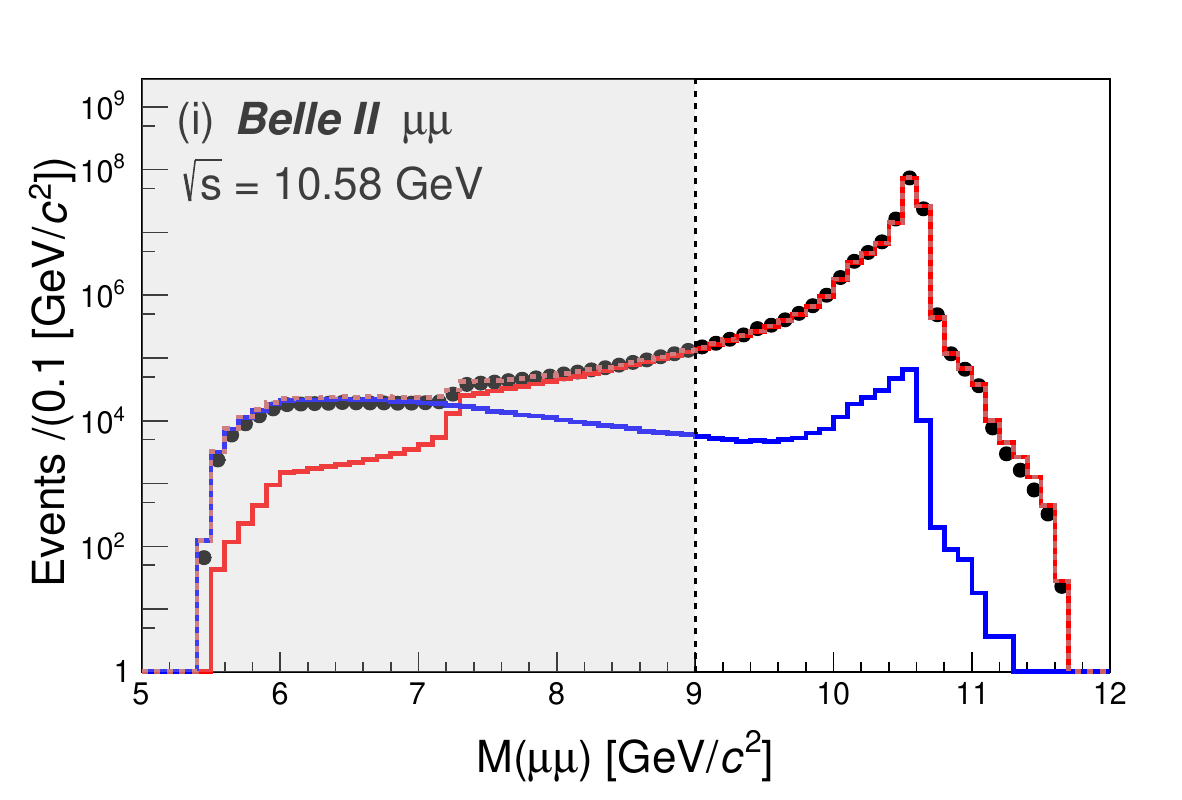}\\
    \end{tabular}
  \caption{Kinematic distributions of dimuon signal candidates in data and MC. The convention in the figure is the same as in Fig.~\ref{fig:dis_bha_4S} except the signal process is dimuon events here. In addition, subfigure (i) shows the invariant mass distribution of the dimuon system (M($\mu\mu$)).}\label{fig:dis_mumu_4S}
\end{figure*}

Signal candidate events are selected following the method outlined below. The digamma candidate events are reconstructed from ECL clusters, while the Bhabha and dimuon candidate events are required to have two tracks with opposite charge. For Bhabha and dimuon events we select the two tracks with the highest momenta, and for digamma events we select the two clusters with the highest energies. We then require that the energies of the selected tracks or clusters fall within the range of 2.50 to 5.82~GeV. Here, energies ($E^{\rm c.m.}$) associated to tracks are determined from the particle momenta measured in the tracking system combined with an assumed mass hypothesis of electron or muon, while the energies of ECL clusters from photon candidates are obtained from the energy depositions in the ECL. To minimize potential systematic uncertainties from trigger (L1 and HLT) efficiency corrections, the selection criteria in the analysis are chosen to be tighter than those in the trigger (details are in Section~\ref{sec:corrections}). We require the energy of the higher-energy particle in the two final-state particles to be greater than 4~GeV ($E_1^{\rm c.m.}>4$~GeV) in Bhabha and digamma candidate events, and greater than 3~GeV ($E_1^{\rm c.m.}>3$~GeV) in dimuon candidate events. We require that the energy deposited in the ECL for a muon candidate track be less than 0.5~GeV. The average ionization energy loss of a muon in the ECL is about 0.2~GeV: therefore a larger energy deposition is likely to be either an electromagnetic shower produced by an electron, or a hadronic interaction of a pion. This criterion effectively rejects Bhabha events from the dimuon processes. To ensure that the two final-state particles are emitted back to back, we require the polar opening angle ($\Delta \theta^{\rm c.m.}_{12} = |\theta_{\rm 1}^{\rm c.m.}+\theta_{\rm 2}^{\rm c.m.} -180\degree|$) to be $\Delta \theta_{\rm 12}^{\rm c.m.}<5\degree$ and the azimuthal opening angle ($\Delta \phi^{\rm c.m.}_{12} = ||\phi^{\rm c.m.}_{\rm 1}-\phi^{\rm c.m.}_{\rm 2}|-180\degree|$) to be $\Delta \phi^{\rm c.m.}_{\rm 12}<5\degree$. We set a tighter requirement for digamma events, $\phi^{\rm c.m.}_{\rm 12}<2.5\degree$. Here, the subscript ``1" (``2'') refers to the final-state particle with the largest (second largest) energy. All clusters must be reconstructed in the ECL with polar angles $37.8\degree<\theta^{\rm cluster}_{\rm 1},~\theta^{\rm cluster}_{\rm 2}<120.5\degree$ in the laboratory frame, which ensures optimal calorimeter energy resolution, avoids detector gaps, and minimizes beam background levels~\cite{axionlike}. To further suppress background contamination in the dimuon measurement, we require the invariant mass of the dimuon system (M($\mu\mu$)) to be greater than 9~GeV$/c^2$. 

The total contaminations of background processes are 0.02\%, 0.66\%, and 0.23\% for Bhabha, digamma, and dimuon channels, respectively. In the digamma channel, a significant contamination of Bhabha events, amounting to (0.55 $\pm$ 0.02)\%, remains even after applying all selection criteria. Thus, we treat both Bhabha and digamma events as signal in the measurement with digamma events, which is hereafter referred to as the digamma-dominated channel. The remaining backgrounds are not accounted for the signal MC samples and considered as a part of the systematic uncertainties in each channel, as described in Section~\ref{sec:sys_uncer}.

The observables used in the selection criteria are shown in Figs.~\ref{fig:dis_bha_4S},~\ref{fig:dis_dig_4S} and \ref{fig:dis_mumu_4S}. In these figures, signal and background MC samples are initially scaled to a common reference luminosity, and then the total MC samples are normalized to the number of events in the data in each figure. The data sample shows good agreement with the MC simulations in most cases, while some discrepancies are observed. In the Bhabha scattering distributions, an excess is observed in $E^{\rm c.m.}_{\rm 1}$ below 4~GeV in Fig.~\ref{fig:dis_bha_4S}(a). This excess is due to the imperfect calibration of the data sample. The discrepancies in the energy distribution have little impact on the luminosity measurement, as the selection criteria require $E^{\rm c.m.}_{\rm 1} >$ 4~GeV. The shapes of the data sample do not match those of the MC simulations in the $\theta^{\rm cluster}_{1}$ and $\theta^{\rm cluster}_{2}$ distributions of Figs.~\ref{fig:dis_bha_4S}, \ref{fig:dis_dig_4S}, and \ref{fig:dis_mumu_4S}. The observed discrepancies suggest a difference in energy reconstruction between data and MC samples. Since the combined $\theta^{\rm cluster}$ distributions exhibit good agreement between the data and the MC, the discrepancies in the $\theta^{\rm cluster}_{1}$ and $\theta^{\rm cluster}_{2}$ distributions have no impact on the luminosity measurement. Discontinuities observed at $\theta^{\rm cluster}=$ 73\degree, 85\degree, 97\degree, and 107\degree~in Figs.~\ref{fig:dis_mumu_4S}(d), \ref{fig:dis_mumu_4S}(e), and \ref{fig:dis_mumu_4S}(f), can be attributed to specific characteristics of the ECL energy cluster reconstruction process. The average ionization energy loss by a muon is approximately 200~MeV, while the minimum value required to form an ECL energy cluster is 30~MeV~\cite{Longo:2020zqt}. Consequently, each muon track typically results in only one ECL cluster associated with a crystal hit. The ECL cluster reconstruction algorithm assigns the position of the cluster to the center of the hit crystal, rather than the actual location of the hit. Additionally, if muons deposit energy at or near the positions of the thin aluminum fins that mechanically support the crystals, there is a higher likelihood of producing single-crystal clusters with energies below the 30~MeV threshold, which are not recorded by the data acquisition system. These two phenomena, along with the choice of histogram bins, result in the observed discontinuities.

After applying all the selection criteria to each signal channel, the selection efficiency for Bhabha channel is found to be (1.88 $\pm$ 0.01)\%. This relatively low efficiency is attributed to: (i) the combined requirement of at least two tracks matched with ECL clusters with energies greater than 1~GeV, which leads to a selection efficiency of (16.27 $\pm$ 0.01)\%; and (ii) for the generated events, limiting electron or positron scattering angles within the detector region of interest, results in a selection efficiency of (16.03 $\pm$ 0.01)\% after the application of the combined requirement in (i). For the digamma-dominated channel, the selection efficiencies are (16.79 $\pm$ 0.01)\% and (0.0016 $\pm$ 0.0001)\% for digamma and Bhabha processes, respectively. The selection efficiency is (29.84 $\pm$ 0.01)\% for the dimuon channel.

\section{Corrections of the L1 trigger and HLT efficiencies}\label{sec:corrections}

The Run~1 data set was collected with an active L1 trigger and HLT. The MC samples are produced with a simulation of the trigger system. The trigger efficiencies are corrected to those measured in data. Trigger efficiencies of the Run~1 data sample are measured by use of the following equation:
\begin{equation}\label{eq:trigger_eff}
  \epsilon_{\rm test} = \frac{N(\text{test} \cap \text{ref})}{N({\rm ref})},
\end{equation}
where ``$N(\text{test} \cap \text{ref})$'' is the number of events that fire both the trigger to be tested and the reference trigger, and ``$N({\rm ref})$'' is the number of events that activate the reference trigger. The test and reference triggers must be independent from each other. Hence, to estimate the L1 trigger efficiency, we choose test triggers solely based on information from the ECL, while reference triggers rely exclusively on CDC information. The test trigger bits selected from the HLT combine information extracted from different subdetectors. We choose reference triggers with looser criteria than those used in the primary analysis so that the reference trigger is 100\% efficient with respect to the analysis selection criteria. These triggers have non-unity prescale factors, which are taken into account. All uncertainties in this section are statistical only.

For the Bhabha channel, events are required to pass an L1 trigger bit, which requires at least one ECL cluster with energy $E^{\rm cluster} > 2~{\rm GeV}$ in the c.m.\ frame falling within the polar angle range of $32.2\degree<\theta^{\rm cluster}<124.6\degree$ in the lab frame. Additionally, after May 2021, a new L1 trigger bit was introduced to the HLT, incorporating a series of criteria to select Bhabha events. In the additional L1 trigger bit, the polar angles and azimuthal angles of two ECL clusters are required to satisfy $|\theta_{1}^{\rm cluster} + \theta_{2}^{\rm cluster}-180\degree|<20\degree$ and $||\phi_{1}^{\rm cluster}-\phi_{2}^{\rm cluster}|-180\degree|<40\degree$ in the c.m.\ frame. The two ECL clusters should both have energies greater than 2.5~GeV, with at least one of them greater than 4~GeV. The L1 trigger efficiencies are calculated to be (99.97 $\pm$ 0.01)\% and (99.69 $\pm$ 0.19)\% for data samples before and after May 2021, respectively. The L1 trigger efficiencies for the MC samples are (98.63 $\pm$ 0.01)\% and (98.15 $\pm$ 0.01)\%. Consequently, correction factors of 1.0136 $\pm$ 0.0001 and 1.0157 $\pm$ 0.0021 are applied to adjust the selection efficiencies of the MC samples for the Bhabha channel. The HLT test trigger requires the electron to fall within the polar angle range of $30\degree<\theta^{\rm cluster}<180\degree$ in the lab frame. The HLT efficiencies are (99.98 $\pm$ 0.01)\% and (99.99 $\pm$ 0.01)\% for the data and MC samples, respectively. The HLT efficiency correction is estimated to be 0.9999 $\pm$ 0.0001 for MC simulation.

For the digamma-dominated channel, the L1 trigger efficiency of the data sample is determined to be (99.97 $\pm$ 0.01)\%. This value is obtained from the Bhabha channel with the L1 trigger bit before May 2021, as both channels utilize the same L1 trigger in this context. The L1 trigger efficiency for the MC samples is (98.59 $\pm$ 0.01)\%. Thus, the correction factor for the L1 trigger is determined to be 1.0140 $\pm$ 0.0001. The HLT bit to select digamma events requires at least one ECL cluster with energy $E^{\rm cluster}>4~{\rm GeV}$. In addition, the energies, polar angles, and azimuthal angles of the two clusters are required to satisfy $2~{\rm GeV}<E_2^{\rm cluster}<E_1^{\rm cluster}$, $32\degree<\theta_1^{\rm cluster}, \theta_2^{\rm cluster}<130\degree$, and $||\phi_{1}^{\rm cluster}-\phi_{2}^{\rm cluster}|-180\degree|<3\degree$. Here, the polar angles are defined in the lab frame, while other quantities are in the c.m.\ frame. The HLT efficiencies for the data and MC samples are calculated to be (99.70 $\pm$ 0.01)\% and (99.99 $\pm$ 0.01)\%, respectively. Consequently, the efficiency correction factor for the HLT is found to be 0.9971 $\pm$ 0.0001.

Dimuon events are selected by an L1 trigger bit that requires the energies, polar angles, and azimuthal angles of the two ECL clusters satisfy $E_{1}^{\rm cluster},E_{2}^{\rm cluster}<2~{\rm GeV}$, $165\degree < \theta^{\rm cluster}_{1}+\theta_{2}^{\rm cluster}<190\degree$, and $160\degree<|\phi_{1}^{\rm cluster}-\phi_{2}^{\rm cluster}|<200\degree$. Due to a misconfiguration in the L1 trigger for data, the calculated L1 trigger efficiencies are (97.33 $\pm$ 0.01)\% for samples collected before August 2020 and (89.48 $\pm$ 0.01)\% for those obtained after. The L1 trigger efficiency for the MC samples is found to be (77.12 $\pm$ 0.01)\%. Thus, the efficiency correction factors for the L1 trigger are determined to be 1.2621 $\pm$ 0.0001 for data samples before August 2020 and 1.1604 $\pm$ 0.0001 for data samples after August 2020. The Belle~II experiment collects data with a redundant set of parallel L1 triggers, however they are not used in this analysis. This redundancy ensures that no physics events were lost due to the misconfiguration issue. The HLT requires momentum thresholds of $p_{1}^{\rm c.m.}>3~{\rm GeV}$ and $p_{2}^{\rm c.m.}>2.5~{\rm GeV}$. Additionally, the energy deposited in the ECL for each muon candidate should be greater than 0~GeV and less than 1~GeV. The HLT efficiencies for the data and MC samples are calculated to be (99.78 $\pm$ 0.01)\% and (99.91 $\pm$ 0.01)\%, respectively. The efficiency correction factor for the HLT is found to be 0.9987 $\pm$ 0.0001.

\section{Determination of integrated luminosities}

Taking into the account the selection efficiencies and their correction factors, the expression for the integrated luminosity becomes 
\begin{equation}
    {\cal L}=\frac{N^{\rm obs}_{\rm data}}{\sigma_{\rm sig}\epsilon_{\rm sig} f_{\rm L1} f_{\rm HLT}}.
\end{equation}
For all three channels, after applying their specific selection criteria, we obtain the numbers of signal events ($N^{\rm obs}_{\rm data}$) in the data sample. The cross sections for the signal processes ($\sigma_{\rm sig}$) are provided by the event generators, while the selection efficiencies of these signal processes ($\epsilon_{\rm sig}$) are estimated using their corresponding MC samples. The correction factors for the L1 trigger and the HLT efficiencies ($f_{\rm L1}$ and $f_{\rm HLT}$) are evaluated as described in Section~\ref{sec:corrections}. All these quantities are listed in Table~\ref{tab:mc_eff}.
\begin{table*}
    \centering
    \caption{Key parameters for the measurement of the integrated luminosity of the $\Upsilon(4S)$ data sample in the three measurement channels. Within the digamma-dominated channel, $\sigma_{\gamma\gamma}$ ($\sigma_{\rm ee}$) signifies the cross section of the digamma (Bhabha) process, while $\epsilon_{\gamma\gamma}$ ($\epsilon_{\rm ee}$) represents the selection efficiency of the digamma (Bhabha) process. Here $f_{\rm L1}^{\rm Tot}$ is the efficiency correction factor for the L1 trigger applied to the entire $\Upsilon(4S)$ sample, which is calculated as a weighted average based on luminosities accumulated during different time periods, as detailed in Section~\ref{sec:corrections} and $f_{\rm HLT}$ is the efficiency correction factor for the HLT. The values in parentheses are the statistical uncertainties, for example, 20.5376 (1) = 20.5376 $\pm$ 0.0001.}\label{tab:mc_eff}
    \renewcommand\arraystretch{1.25}
    \begin{tabular}{C{2cm} C{2cm} C{3cm} C{3cm} C{2cm} C{2cm} C{2cm}}
    \hline\hline
    Channel & $N_{\rm data}^{\rm obs}$ $(\times 10^8)$ & $\sigma_{\rm sig}$ (nb) & $\epsilon_{\rm sig}$ (\%) & $f_{\rm L1}^{\rm Tot}$ & $f_{\rm HLT}$ & $\cal L$ (fb$^{-1}$)\\\hline
    Bhabha & 20.5376 (1) & {295.38 (4)} & 1.88 (1) & 1.0148 (17) & 0.9999 (1) & {364.48} (3)\\
    Digamma- & \multirow{2}{*}{3.1228 (2)} & $\sigma_{\gamma\gamma}$: {5.0686 (5)} & $\epsilon_{\gamma\gamma}$: 16.53 (1) &\multirow{2}{*}{1.0140 (1)} & \multirow{2}{*}{0.9971 (1)} & \multirow{2}{*}{{366.57} (3)}\\
    dominated & & $\sigma_{\rm ee}$: {295.38 (4)} & $\epsilon_{\rm ee}$: 0.16 (1) $\times 10^{-2}$ & & &\\
    Dimuon & 1.2344 (1) & {1.1472 (1)} & 25.40 (1) & 1.1719 (1) & 0.9987 (1) & {361.97} (4)\\
    \hline\hline
    \end{tabular}
\end{table*}

The integrated luminosity of the $\Upsilon(4S)$ data sample is determined to be ({364.48} $\pm$ 0.03)~fb$^{-1}$, ({366.57} $\pm$ 0.03)~fb$^{-1}$, and ({361.97} $\pm$ 0.04)~fb$^{-1}$ with the Bhabha, digamma-dominated, and dimuon channels, respectively. The uncertainties are statistical only and obtained from data. The uncertainties due to the limited sizes of the MC samples are treated as a part of the systematic uncertainties in this paper.

\section{Systematic uncertainties}\label{sec:sys_uncer}

\begin{table}
  \centering
  \caption{Contributions to the systematic uncertainties of the measured integrated luminosities at $\Upsilon(4S)$ for the Bhabha, digamma-dominated, and dimuon channels. The uncertainties denoted with a superscript * represent estimations based on time-independent MC samples, while the others are based on time-dependent samples.}\label{tab:uncer}
  \renewcommand\arraystretch{1.25}
  \begin{tabular}{c C{1.5cm} C{1.5cm} C{1.5cm}}
    \hline\hline
    Source & $e^+e^-$(\%) & $\gamma\gamma$(\%) & $\mu^{+}\mu^{-}$(\%)\\\hline
    Cross section & {{$\pm0.23$}} & {$\pm0.15$} & $\pm0.44$\\
    $\sqrt{s}$ (c.m.\ energy)* & $\pm0.15$& $\pm0.25$ & $\pm0.29$\\
    Input angular range* & $\pm0.08$& $\pm0.01$& $\pm0.03$\\
    ECL alignment* & $\pm0.02$ & $\pm0.02$& $\pm0.02$\\
    MC statistics & $\pm0.03$& $\pm0.02$ & $\pm0.02$\\
    Beam background* & $\pm0.13$ & $\pm0.19$ &$\pm0.26$ \\
    Track reconstruction & $\pm0.48$ & --- & $\pm0.48$\\
    Cluster reconstruction & --- & $\pm0.41$ & ---\\
    Charge misassignment & $\pm0.03$ & --- & $\pm0.01$\\
    $E^{\rm c.m.}$ criteria & $\pm0.04$ & $\pm0.07$ & $\pm0.09$\\
    $\theta^{\rm cluster}$ criteria & $\pm0.08$& $\pm0.06$ & $\pm0.11$\\
    $\Delta\theta^{\rm c.m.}_{12}$ criteria & $\pm0.05$ & $\pm0.07$ & $\pm0.29$\\
    $\Delta\phi^{\rm c.m.}_{12}$ criteria & $\pm0.01$& $\pm0.04$& $\pm0.26$\\
    Cluster veto criteria & --- & --- & $\pm0.03$\\
    Material effects* & $\pm0.05$& $\pm0.20$ & ---\\
    Overlapping clusters & --- & $\pm0.01$ & ---\\
    Background processes& $\pm0.02$ & $\pm0.11$& $\pm0.23$\\
    L1Trigger & $\pm0.16$& $\pm0.03$& $\pm0.01$\\\hline
    Quadrature sum & {$\pm0.61$} & {$\pm0.60$} & $\pm0.90$\\
    \hline\hline
  \end{tabular}
\end{table}

Table~\ref{tab:uncer} lists eighteen contributions to the systematic uncertainties. Here, we provide a complete description of the systematic uncertainties for the $\Upsilon(4S)$ sample, and the procedure is the same for the other samples. We find small changes in the calculated luminosity when the selection criteria or the generation parameters within MC samples are adjusted.
The difference between the modified luminosity and the nominal luminosity is taken as the systematic uncertainty, unless otherwise stated. 

The BABAYAGA@NLO generator provides {a theoretical cross-section uncertainty of 0.20\% for the Bhabha process and 0.10\% for the digamma process}~\cite{Bhabha-gen, digamma-gen}. {A small fraction of generated events exceeds the phase space, impacting the cross sections by 0.10\%. Thus, the total cross-section uncertainties for Bhabha and digamma processes are conservatively estimated to be {0.23\%} and 0.15\%, respectively.} The precision of the dimuon process is computed by the KKMC generator as an uncertainty of 0.44\%~\cite{Jadach:1999vf,Banerjee:2007is}.

To assess the impact of the input parameters of the generators, we vary each parameter and measure the luminosity difference relative to the nominal value, treating each change as a systematic uncertainty associated with that parameter. The c.m.\ energy fluctuates around the $\Upsilon(4S)$ peak with less than a 5~MeV variation over time. To cover the impact of this potential deviation, we vary the c.m.\ energy, increasing or decreasing it by 5~MeV in the signal MC samples. These adjustments result in uncertainties of 0.15\%, 0.25\%, and 0.29\% for the Bhabha, digamma-dominated, and dimuon channels, respectively.

The angular range of the primary final-state particles is an important input for generators. To improve computational efficiency, low multiplicity events are simulated only if the final-state particles are expected to interact in the active volume of the detector. To estimate boundary effects from scattering, we change the polar angle range to $35^\circ$--$145^\circ$. The uncertainties are estimated as 0.08\%, 0.01\%, and 0.03\% for the Bhabha, digamma-dominated, and dimuon channels, respectively.

Following the alignment of ECL crystal positions using dimuon events, uncertainties of 0.04~mm in the $x$ direction, 0.08~mm in the $y$ direction, and 0.11~mm in the $z$ direction have been determined. These uncertainties are equivalent to uncertainty in the position of the IP. We generate a series of new MC samples for the three channels, incorporating this uncertainty in the IP position. The uncertainties related to the alignment of the ECL location are estimated to be approximately 0.02\% for the Bhabha, digamma-dominated, and dimuon channels.

The limited MC sample sizes lead to uncertainties in the measurement of the selection efficiencies. The uncertainties are determined to be 0.03\% for the Bhabha channel and 0.02\% for digamma-dominated and dimuon channels, respectively.

As the instantaneous luminosity increases, the impact of beam background becomes a larger component of the systematic uncertainty. To assess this uncertainty, we calculate the weighted average of the differences between the luminosities computed using time-dependent MC samples, which includes realistic beam background, and time-independent MC samples characterized by significantly lower beam background, calculated from simulation. Since the time-dependent MC samples describe the data well, we consider half of the differences as the systematic uncertainties associated with beam background, amounting to 0.13\%, 0.19\%, and 0.26\% for the Bhabha, digamma-dominated, and dimuon channels, respectively.

Uncertainties in track reconstruction efficiencies arise during the track reconstruction procedure, where the difference between the data and the MC simulations may reach 0.24\% per track~\cite{Bertacchi:2020eez}. As Bhabha and dimuon events each have two tracks, and assuming a 100\% correlation between them, we obtain a conservative uncertainty of 0.48\% for the Bhabha and dimuon channels. The uncertainty of the cluster reconstruction for the digamma-dominated channel is estimated to be 0.41\%. A measurement without any requirements on the cluster finds that the uncertainty associated with cluster reconstruction is negligible for the Bhabha channel.

Two tracks with opposite charges are required in the Bhabha and dimuon channels. To determine the impact of potential charge misassignment, we perform new measurements by removing the criterion related to charge and obtain uncertainties of 0.03\% and 0.01\% for the measurements with Bhabha and dimuon events, respectively.

The criteria for selecting signal events may introduce uncertainties. We vary the selection requirements by tightening or loosening them as follows:

\vspace{10pt}
\textbullet~$2.50^{+0.25}_{-0.25}$~GeV $<E^{\rm c.m.}_{\rm 2}<E^{\rm c.m.}_{\rm 1}<$ $5.82^{+0.25}_{-0.25}$~GeV;\vspace{10pt}

\textbullet~$ {37.8\degree}^{+1.6\degree}_{-2.8\degree} < \theta^{\rm cluster}_{\rm 1}, ~\theta^{\rm cluster}_{\rm 2} < {120.5\degree}^{+4.1\degree}_{-2.1\degree}$;\vspace{10pt}

\textbullet~$\Delta\theta^{\rm c.m.}_{\rm 12}<{5\degree}^{+2.5\degree}_{-2.5\degree}$;\vspace{10pt}

\textbullet~$\Delta\phi^{\rm c.m.}_{\rm 12}<{5\degree}^{+2.5\degree}_{-2.5\degree}$;\vspace{10pt}

\textbullet~$\Delta\phi^{\rm c.m.}_{\rm 12}<{2.5\degree}^{+1.0\degree}_{-1.0\degree}$ (Only for digamma-dominated channel);\vspace{10pt}

\textbullet~$0<E^{\rm cluster}_{\rm 2},~E^{\rm cluster}_{\rm 1}<0.50^{+0.02}_{-0.02}$~GeV (cluster veto criteria for dimuon channel).\vspace{10pt}

\noindent The energy resolution of the final-state particles at an energy of about 5.29~GeV is determined to be approximately 0.08~GeV. To estimate the systematic uncertainty arising from the $E^{\rm c.m.}$ criteria, we roughly triple the energy resolution. This method is also used to estimate the uncertainty associated with cluster veto criteria. For the $\theta^{\rm cluster}$ criteria, we change the polar angle range by adding or removing ECL crystal rings. The adjustments to the requirements on $\Delta\theta_{\rm 12}^{\rm c.m.}$ and $\Delta\phi_{\rm 12}^{\rm c.m.}$ are approximately expanded or reduced by half from the original range. When we change each selection criterion, the larger difference in integrated luminosity compared to our standard result is taken as the systematic uncertainty. The systematic uncertainties from all sources are tabulated in Table~\ref{tab:uncer}.

Photons, electrons, and positrons may interact with the material in the VXD, which causes the production and absorption of photons, electrons and positrons. To investigate the size of this effect, we remove the simulation of the VXD material in MC samples. The uncertainties are approximately 0.05\% for the Bhabha channel and 0.20\% for the digamma-dominated channel.

After a photon interacts with the detector material, it may convert into two nearby clusters in the ECL. Selecting only one cluster per photon introduces bias. To address this issue, we employ additional criteria to treat the cluster with the smallest opening angle within a 5\degree~range in both polar and azimuthal angles relative to each selected cluster as a single overlapping cluster. The difference in luminosity, including the presence or absence of overlapping clusters, serves as a measure of systematic uncertainty. This uncertainty of approximately 0.01\% is specific to the digamma-dominated channel, since the energies of Bhabha and dimuon tracks are obtained in a different manner. 

As described in Section~\ref{sec:data_mc}, various background processes are produced and analyzed to assess their impact on the luminosity measurement. The primary background channels (background levels) are digamma ($9.3 \times 10^{-5}$), $u{\bar u}$ ($8.3\times10^{-4}$), and $e^+e^-\mu^+\mu^-$ ($2.0\times10^{-3}$) events for the measurements with Bhabha, digamma-dominated, and dimuon channels, respectively. Given that the size of the contributions are relatively low, we assign a 100\% uncertainty on the contribution from these background processes. Consequently, the total uncertainties due to background processes are 0.02\%, 0.11\%, and 0.23\% for these three channels, respectively.

Uncertainties arising from the trigger efficiency correction procedures are due to the limited sizes of the samples with the reference trigger applied. The uncertainties are measured to be 0.16\%, 0.03\%, and 0.01\% for the Bhabha, digamma-dominated, and dimuon channels for the L1 trigger efficiency corrections, respectively. These statistical uncertainties are treated as the systematic uncertainties corresponding to the relative trigger bits.

The combined relative uncertainties are calculated by adding the individual uncertainties in quadrature, assuming that they are uncorrelated. The resulting total relative uncertainties of the $\Upsilon(4S)$ sample are {0.61\%}, 0.60\%, and 0.90\% for the Bhabha, digamma-dominated, and dimuon channels, respectively. Most of the systematic uncertainties are positively correlated for the $\Upsilon(4S)$ and off-$\Upsilon(4S)$ samples, except for those associated with ``c.m.\ energy", ``MC statistics", ``beam background", and ``$E^{\rm c.m.}$ criteria". The Run~1 luminosity is measured to be ({426.88} $\pm$ 0.03 $\pm$ {2.61})~fb$^{-1}$, ({429.28} $\pm$ 0.03 $\pm$ {2.62})~fb$^{-1}$, and ({423.99} $\pm$ 0.04 $\pm$ {3.83})~fb$^{-1}$ with the Bhabha, digamma-dominated, and dimuon channels, respectively.

The integrated luminosities of the data sample at various energy points are given in Table~\ref{tab:lum}. The measured luminosities with Bhabha, digamma-dominated, and dimuon channels are consistent within {1.3$\sigma$} after the removal of correlated uncertainties. The luminosities obtained with the three channels are combined by considering the correlation of the uncertainties among them. The mean values ($\cal L$) and the uncertainties ($\Delta {\cal L}$) are calculated with~\cite{DAgostini:1993arp}
\begin{equation}
    {\bar {\cal L}} \pm \Delta {\cal L} = \frac{\Sigma_i {\cal L}_i \cdot \Sigma_j \omega_{ij}}{\Sigma_i \Sigma_j \omega_{ij}} \pm \sqrt{\frac{1}{\Sigma_i \Sigma_j \omega_{ij}}},
\end{equation}
where $i$ and $j$ are summed over Bhabha, digamma-dominated, and dimuon channels, $w_{ij}$ is the element of the weight matrix $W = V^{-1}$, and $V$ is the covariance error matrix calculated according to the statistical and systematic uncertainties in Table~\ref{tab:lum}. Combining the results of all three channels, the error matrix can be calculated as
\begin{equation}
V=
\begin{pmatrix}
    \Delta {\cal L}_{\rm ee}^2 & {\cal L}_{\rm ee} {\cal L}_{\rm \gamma \gamma} \delta_{\rm ee \gamma \gamma}^2 & {\cal L}_{\rm ee} {\cal L}_{\rm \mu \mu} \delta_{\rm ee \mu \mu}^2\\
    {\cal L}_{\rm ee} {\cal L}_{\rm \gamma \gamma} \delta_{\rm ee \gamma \gamma}^2 & \Delta {\cal L}_{\rm \gamma \gamma}^2 & {\cal L}_{\rm \gamma \gamma} {\cal L}_{\rm \mu \mu} \delta_{\rm \gamma \gamma \mu \mu}^2\\
    {\cal L}_{\rm ee} {\cal L}_{\rm \mu \mu} \delta_{\rm ee \mu \mu}^2 & {\cal L}_{\rm \gamma \gamma} {\cal L}_{\rm \mu \mu} \delta_{\rm \gamma \gamma \mu \mu}^2 & \Delta {\cal L}_{\rm \mu \mu}^2\\
\end{pmatrix},
\end{equation}
where the luminosity $\cal {\cal L}$ (total uncertainty $\Delta {\cal L}$) with subscripts ${\rm ee}$, ${\rm \gamma \gamma}$, or ${\rm \mu \mu}$ represent the luminosity (total uncertainty) is obtained from Bhabha, digamma-dominated, or dimuon channels. The symbol $\delta$ with a combination of two subscripts denotes the common relative systematic uncertainties between two corresponding channels. Here, $\delta_{\rm ee \gamma \gamma}={0.26}$\%, $\delta_{\rm ee \mu \mu}={0.52}$\%, and $\delta_{\rm \gamma \gamma \mu \mu}=0.32$\%. The results from this averaging procedure are given in Table~\ref{tab:lum}. The total uncertainty of the average luminosity is about {0.47\%}.

\begin{table*}[htbp]
  \centering
  \caption{The integrated luminosities of the data sample at different energy points. The quantities ${\cal L}_{\rm ee}$, ${\cal L}_{\rm \gamma \gamma}$, and ${\cal L}_{\rm \mu \mu}$ are the integrated luminosities obtained with the Bhabha, digamma-dominated, and dimuon channels, respectively. In the last column $\cal L$ denotes the combined results of the three luminosity measurement channels. The first uncertainties represent statistical uncertainties, while the second are systematic uncertainties. The uncertainties of combined luminosities are the total uncertainties, which include both statistical and systematic uncertainties.}\label{tab:lum}
  \vspace{0.2cm}
  \renewcommand\arraystretch{1.25}
  \begin{tabular}{C{3.2cm} C{2cm} C{3.2cm} C{3.2cm} C{3.2cm} C{2.2cm}}
  \hline\hline
Type & $\sqrt{s}$ (GeV) & ${\cal L}_{\rm ee}$ (fb$^{-1}$) & ${\cal L}_{\rm \gamma\gamma}$ (fb$^{-1}$) & ${\cal L}_{\rm \mu\mu}$ (fb$^{-1}$) & $\cal L$ (fb$^{-1}$)\\\hline
$\Upsilon(4S)$ & 10.580 & {364.48} $\pm$ 0.03 $\pm$ {2.23} & {366.57} $\pm$ 0.03 $\pm$ {2.20} & {361.97} $\pm$ 0.04 $\pm$ {3.26} & {365.37 $\pm$ {1.70}}\\
off-$\Upsilon(4S)$ & {10.517} & {42.60} $\pm$ 0.01 $\pm$ 0.26 & {42.90} $\pm$ 0.01 $\pm$ 0.26 & {42.41} $\pm$ 0.02 $\pm$ 0.39 & {42.74} $\pm$ 0.20\\
\hline
\multirow{4}{*}{$\Upsilon(5S)$ scan}&10,657& 3.55 $\pm$ 0.01 $\pm$ 0.03 & {3.55} $\pm$ 0.01 $\pm$ 0.03 & 3.51 $\pm$ 0.01 $\pm$ 0.04 & 3.54 $\pm$ 0.03\\
&10.706 & 1.64 $\pm$ 0.01 $\pm$ 0.02 & 1.64 $\pm$ 0.01 $\pm$ 0.02 & 1.62 $\pm$ 0.01 $\pm$ 0.02 & 1.63 $\pm$ 0.02\\
&10.751 & {9.88} $\pm$ 0.01 $\pm$ 0.07 & {9.91} $\pm$ 0.01 $\pm$ 0.08 & 9.78 $\pm$ 0.01 $\pm$ 0.10 & {9.88} $\pm$ 0.06\\
&10.810 & {4.72} $\pm$ 0.01 $\pm$ 0.04 & {4.71} $\pm$ 0.01 $\pm$ 0.04 & 4.69 $\pm$ 0.01 $\pm$ 0.05 & {4.71} $\pm$ 0.03\\
\hline
Total & --- & {426.88} $\pm$ 0.03 $\pm$ {2.61} & {429.28} $\pm$ 0.03 $\pm$ {2.62} & {423.99} $\pm$ 0.04 $\pm$ {3.83} & {427.87 $\pm$ 2.01}\\
\hline\hline
\end{tabular}
\end{table*}

\section{Conclusions}

The integrated luminosity of the Run~1 data sample collected from March 2019 to June 2022 with the Belle~II detector at SuperKEKB is measured with Bhabha, digamma, and dimuon events. We determine the total integrated luminosity of the data sample to be ({426.88} $\pm$ 0.03 $\pm$ {2.61})~fb$^{-1}$, ({429.28} $\pm$ 0.03 $\pm$ {2.62})~fb$^{-1}$, and ({423.99} $\pm$ 0.04 $\pm$ {3.83})~fb$^{-1}$ with the Bhabha, digamma-dominated, dimuon channels, where the first uncertainties are statistical and the second are systematic. The luminosity results obtained from three channels are consistent within 1$\sigma$. The combined luminosity obtained from these three luminosity results is ({427.87 $\pm$ 2.01})~fb$^{-1}$. The integrated luminosities of individual data samples at different energies are summarized in Table~\ref{tab:lum}. These integrated luminosities, measured in this work, serve as basic inputs for many analyses at Belle~II.

\section{Acknowledgements}
This work, based on data collected using the Belle II detector, which was built and commissioned prior to March 2019,
was supported by
Higher Education and Science Committee of the Republic of Armenia Grant No.~23LCG-1C011;
Australian Research Council and Research Grants
No.~DP200101792, 
No.~DP210101900, 
No.~DP210102831, 
No.~DE220100462, 
No.~LE210100098, 
and
No.~LE230100085; 
Austrian Federal Ministry of Education, Science and Research,
Austrian Science Fund
No.~P~34529,
No.~J~4731,
No.~J~4625,
and
No.~M~3153,
and
Horizon 2020 ERC Starting Grant No.~947006 ``InterLeptons'';
Natural Sciences and Engineering Research Council of Canada, Compute Canada and CANARIE;
National Key R\&D Program of China under Contract No.~2022YFA1601903,
National Natural Science Foundation of China and Research Grants
No.~11575017,
No.~11761141009,
No.~11705209,
No.~11975076,
No.~12135005,
No.~12150004,
No.~12161141008,
and
No.~12175041,
and Shandong Provincial Natural Science Foundation Project~ZR2022JQ02;
the Czech Science Foundation Grant No.~22-18469S 
and
Charles University Grant Agency project No.~246122;
European Research Council, Seventh Framework PIEF-GA-2013-622527,
Horizon 2020 ERC-Advanced Grants No.~267104 and No.~884719,
Horizon 2020 ERC-Consolidator Grant No.~819127,
Horizon 2020 Marie Sklodowska-Curie Grant Agreement No.~700525 ``NIOBE''
and
No.~101026516,
and
Horizon 2020 Marie Sklodowska-Curie RISE project JENNIFER2 Grant Agreement No.~822070 (European grants);
L'Institut National de Physique Nucl\'{e}aire et de Physique des Particules (IN2P3) du CNRS
and
L'Agence Nationale de la Recherche (ANR) under grant ANR-21-CE31-0009 (France);
BMBF, DFG, HGF, MPG, and AvH Foundation (Germany);
Department of Atomic Energy under Project Identification No.~RTI 4002,
Department of Science and Technology,
and
UPES SEED funding programs
No.~UPES/R\&D-SEED-INFRA/17052023/01 and
No.~UPES/R\&D-SOE/20062022/06 (India);
Israel Science Foundation Grant No.~2476/17,
U.S.-Israel Binational Science Foundation Grant No.~2016113, and
Israel Ministry of Science Grant No.~3-16543;
Istituto Nazionale di Fisica Nucleare and the Research Grants BELLE2;
Japan Society for the Promotion of Science, Grant-in-Aid for Scientific Research Grants
No.~16H03968,
No.~16H03993,
No.~16H06492,
No.~16K05323,
No.~17H01133,
No.~17H05405,
No.~18K03621,
No.~18H03710,
No.~18H05226,
No.~19H00682, 
No.~20H05850,
No.~20H05858,
No.~22H00144,
No.~22K14056,
No.~22K21347,
No.~23H05433,
No.~26220706,
and
No.~26400255,
and
the Ministry of Education, Culture, Sports, Science, and Technology (MEXT) of Japan;  
National Research Foundation (NRF) of Korea Grants
No.~2016R1\-D1A1B\-02012900,
No.~2018R1\-A2B\-3003643,
No.~2018R1\-A6A1A\-06024970,
No.~2019R1\-I1A3A\-01058933,
No.~2021R1\-A6A1A\-03043957,
No.~2021R1\-F1A\-1060423,
No.~2021R1\-F1A\-1064008,
No.~2022R1\-A2C\-1003993,
and
No.~RS-2022-00197659,
Radiation Science Research Institute,
Foreign Large-Size Research Facility Application Supporting project,
the Global Science Experimental Data Hub Center of the Korea Institute of Science and Technology Information
and
KREONET/GLORIAD;
Universiti Malaya RU grant, Akademi Sains Malaysia, and Ministry of Education Malaysia;
Frontiers of Science Program Contracts
No.~FOINS-296,
No.~CB-221329,
No.~CB-236394,
No.~CB-254409,
and
No.~CB-180023, and SEP-CINVESTAV Research Grant No.~237 (Mexico);
the Polish Ministry of Science and Higher Education and the National Science Center;
the Ministry of Science and Higher Education of the Russian Federation
and
the HSE University Basic Research Program, Moscow;
University of Tabuk Research Grants
No.~S-0256-1438 and No.~S-0280-1439 (Saudi Arabia);
Slovenian Research Agency and Research Grants
No.~J1-9124
and
No.~P1-0135;
Agencia Estatal de Investigacion, Spain
Grant No.~RYC2020-029875-I
and
Generalitat Valenciana, Spain
Grant No.~CIDEGENT/2018/020;
The Knut and Alice Wallenberg Foundation (Sweden), Contracts No.~2021.0174 and No.~2021.0299;
National Science and Technology Council,
and
Ministry of Education (Taiwan);
Thailand Center of Excellence in Physics;
TUBITAK ULAKBIM (Turkey);
National Research Foundation of Ukraine, Project No.~2020.02/0257,
and
Ministry of Education and Science of Ukraine;
the U.S. National Science Foundation and Research Grants
No.~PHY-1913789 
and
No.~PHY-2111604, 
and the U.S. Department of Energy and Research Awards
No.~DE-AC06-76RLO1830, 
No.~DE-SC0007983, 
No.~DE-SC0009824, 
No.~DE-SC0009973, 
No.~DE-SC0010007, 
No.~DE-SC0010073, 
No.~DE-SC0010118, 
No.~DE-SC0010504, 
No.~DE-SC0011784, 
No.~DE-SC0012704, 
No.~DE-SC0019230, 
No.~DE-SC0021274, 
No.~DE-SC0021616, 
No.~DE-SC0022350, 
No.~DE-SC0023470; 
and
the Vietnam Academy of Science and Technology (VAST) under Grants
No.~NVCC.05.12/22-23
and
No.~DL0000.02/24-25.

These acknowledgements are not to be interpreted as an endorsement of any statement made
by any of our institutes, funding agencies, governments, or their representatives.

We thank the SuperKEKB team for delivering high-luminosity collisions;
the KEK cryogenics group for the efficient operation of the detector solenoid magnet and IBBelle on site;
the KEK Computer Research Center for on-site computing support; the NII for SINET6 network support;
and the raw-data centers hosted by BNL, DESY, GridKa, IN2P3, INFN, 
and the University of Victoria.

\bibliography{references}

\end{document}